\documentclass[a4paper]{amsart}
\usepackage{amsmath, amsthm, amssymb, amscd, mathrsfs, eucal,
  amsfonts, latexsym}

\newcommand{\rest}{\upharpoonright}
\newcommand{\calA}{{\mathcal A}}
\newcommand{\tildA}{\tilde{\mathcal A}}
\newcommand{\tildAnul}{\tilde{\mathcal A}_0}
\newcommand{\al}{\alpha}
\newcommand{\calB}{{\mathcal B}}
\newcommand{\be}{\beta}
\newcommand{\calC}{{\mathcal C}}
\newcommand{\calD}{{\mathcal D}}
\newcommand{\Co}{{\mathbb C}}
\newcommand{\cO}{{\mathcal O}}                                      
\newcommand{\cC}{{\mathcal C}}                                      
\newcommand{\de}{\delta}
\newcommand{\De}{\Delta}
\newcommand{\calE}{{\mathcal E}}
\newcommand{\calF}{{\mathcal F}}
\newcommand{\calG}{{\mathcal G}}
\newcommand{\f}{\varphi}
\newcommand{\ga}{\gamma}
\newcommand{\calH}{{\mathcal H}}
\newcommand{\calK}{{\mathcal K}}
\newcommand{\calS}{{\mathcal S}}
\newcommand{\frakC}{{\mathfrak C}}
\newcommand{\frakD}{{\mathfrak D}}
\newcommand{\frakK}{{\mathfrak K}}
\newcommand{\frakM}{{\mathfrak{Man}}}
\newcommand{\frakA}{{\mathfrak{Alg}}}
\newcommand{\frakTa}{{\mathfrak{TAlg}}}
\newcommand{\frakR}{{\mathfrak{Rep}}}
\newcommand{\frakS}{{\mathfrak{Sts}}}
\newcommand{\fraktest}{{\mathfrak{Test}}}
\newcommand{\fraktopvec}{{\mathfrak{Top}}}
\newcommand{\euM}{{\mathscr M}}
\newcommand{\euN}{{\mathscr N}} 
\newcommand{\euA}{{\mathscr A}}
\newcommand{\euW}{{\mathscr W}}
\newcommand{\euD}{{\mathscr D}}
\newcommand{\La}{\Lambda}
\newcommand{\calL}{{\mathcal L}}
\newcommand{\calM}{{\mathcal M}}
\newcommand{\calN}{{\mathcal N}}
\newcommand{\m}{\mu}
\newcommand{\Na}{{\mathbb N}}
\newcommand{\om}{\omega}
\newcommand{\calO}{{\mathcal O}}
\newcommand{\p}{\pi}
\newcommand{\Q}{\Omega}
\newcommand{\calR}{{\mathcal R}}
\newcommand{\R}{{\mathbb R}}
\newcommand{\si}{\sigma}
\newcommand{\ta}{\tau}
\newcommand{\Th}{\Theta}
\newcommand{\calU}{{\mathcal U}}
\newcommand{\calW}{{\mathcal W}}
\newcommand{\x}{\xi}
\newcommand{\calZ}{{\mathcal Z}}
\newcommand{\Ze}{{\mathbb Z}}
\newcommand{\ad}{{\rm ad}}
\newcommand{\Poi}{{\mathcal P}}
\newcommand{\Poif}{{\mathcal P}_+^\uparrow}
\newcommand{\Poip}{{\mathcal P}_+^\downarrow}
\newcommand{\ProPoi}{{\mathcal P}_+}
\newcommand{\slduer}{\mbox{SL}(2,{\R})} 
\newcommand{\Sn}{\widetilde{\mathcal P}}
\newcommand{\imply}{\Rightarrow}
\newcommand{\np}{\par\noindent}
\newcommand{\lar}{\longrightarrow}
\newcommand{\spann}{\mbox{span}}
\newcommand{\supp}{\mbox{supp}}
\newcommand{\WF}{\mbox{\rm WF}}
\newcommand{\degree}{\mbox{deg}}
\newcommand{\sd}{\mbox{sd}}
\newcommand{\obj}{\mbox{\rm Obj}}
\newcommand{\Hom}{\mbox{\rm hom}}
\newcommand{\trace}{\mbox{tr}}
\newcommand{\Lor}{\mbox{\tiny Lor}}
\newcommand{\reg}{\mbox{\tiny reg}}
\newcommand{\unit}{{\mbox{id}}}
\newcommand{\bC}{{\mathbb C}}
\newcommand{\gb}{\boldsymbol{g}}
\newcommand{\gbo}{\overset{\circ}{{\boldsymbol g}}}
\newcommand{\nnn}{\circ}
\newcommand{\calRo}{\overset{\circ}{\mathcal{R}}}
\newcommand{\sigmao}{\overset{\circ}{\sigma}}
\newcommand{\Eo}{\overset{\circ}{E}}
\newcommand{\Ko}{\overset{\circ}{K}}
\newcommand{\Wo}{\overset{\circ}{W}}
\newcommand{\Fo}{\Phi_{\omega}}
\newcommand{\Fco}{\check{\Phi}_{\omega}}
\newcommand{\ggo}{\overset{\circ}{g}}

\newcommand{\triv}{\boldsymbol{0}}
\theoremstyle{plain}
\swapnumbers
\newtheorem{Thm}{Theorem}[section]
\newtheorem{Pro}[Thm]{Proposition}

\newtheorem{Def}[Thm]{Definition}
\theoremstyle{remark}

\begin{document}

\title[The Generally Covariant Locality Principle]{}
\begin{center}
{\Large \bf The Generally
  Covariant Locality Principle \\[6pt] --
 A New Paradigm for Local Quantum Field Theory}
${}$\\[16pt]
\author[Brunetti, Fredenhagen and Verch]{}
{\sc Romeo Brunetti}$^{(1)}$, 
 {\sc Klaus 
Fredenhagen}$^{(2)}$ {\rm and} {\sc Rainer Verch}$^{(3)}$\\
\hfill\\
\hfill\\
{\small $^{(1)}$ Dip.\ Scienze Fisiche, Univ.\ Napoli ``Federico II",
 Comp.\ Univ.\ 
Monte Sant'Angelo, Via Cintia, I-80126 Napoli,  and I.N.F.N. sez. 
Napoli, Italy }\\[4pt]
{\small $^{(2)}$ II.\ Inst.\ f.\ Theoretische Physik, Universit\"at
  Hamburg,\\ Luruper  
Chaussee 149, D-22761 Hamburg, Germany}\\[4pt]
{\small $^{(3)}$ Inst.\ f.\ Theoretische Physik, Universit\"at G\"ottingen, \\
Bunsenstrasse 9, D-37073 G\"ottingen, Germany}\\
\hfill\\
{\small {e-mail: \tt brunetti@na.infn.it, fredenha@x4u2.desy.de, 
verch@theorie.physik.uni-goettingen.de}
}
\end{center}

\begin{abstract}
A new approach to the model-independent description of quantum field
theories will be introduced in the present work. The main feature of
this new approach is to incorporate in a local sense the principle of
general covariance of general relativity, thus giving rise to the
concept of a {\it locally covariant quantum field theory}.
Such locally covariant quantum field theories will be described
mathematically  in terms of 
covariant functors between the categories, on one side, of globally 
hyperbolic spacetimes with isometric embeddings as morphisms and, on the 
other side, of $^\ast$-algebras with unital injective 
$^\ast$-endomorphisms as morphisms. Moreover, locally covariant
quantum fields can be described in this framework as natural
transformations between certain functors. The usual Haag-Kastler
framework of nets of operator-algebras over a fixed
spacetime background-manifold, together with covariant automorphic actions of
the isometry-group of the  background spacetime, can be re-gained from
this new approach as a special case.
 Examples of this new approach are also outlined. In
case that a locally covariant quantum field theory obeys the
time-slice axiom, one can naturally associate to it certain
automorphic actions, 
called ``relative Cauchy-evolutions'', which describe the dynamical
reaction of the quantum field theory to a local change of spacetime background
metrics. The functional derivative of a relative Cauchy-evolution
with respect to the spacetime metric is found to be a divergence-free
quantity which has, as will be demonstrated in an example, the
significance of an energy-momentum tensor for the locally covariant 
quantum field theory. Furthermore, we discuss the functorial
properties of state spaces of locally covariant quantum field theories
that entail the validity of the principle of local definiteness.
\end{abstract}
\maketitle

\section{Introduction}
Quantum field theory incorporates two main principles into quantum
physics, locality and covariance. Locality expresses the idea that
quantum processes can be localized in space and time [and, at the
level of observable quantities, that causally separated processes are
exempt from any uncertainty relations restricting their
commensurability]. The principle of Poincar\'e-covariance within
special relativity states that there are no preferred Lorentzian
coordinates for the description of physical processes, and thereby the
concept of an absolute space as an arena for physical phenomena is
abandoned. Yet it is still meaningful to speak of events in terms of
spacetime points as entities of a given, fixed spacetime background in
the setting of special relativistic physics.

In general relativity, however, spacetime points lose this a priori
meaning. The principle of general covariance forces one to regard
spacetime points simultaneously as members of several, locally
diffeomorphic spacetimes. It is rather the relations between
distinguished events that have a physical interpretation. 

This principle should also be observed when quantum field theory in the
presence of gravitational fields is discussed. A first approximation to such
situations is to consider quantum fields on a given, curved
Lorentzian background spacetime
where the sources of the gravitational curvature are
described classically and independently of the dynamics of the quantum
fields in that background. Due to the weakness of gravitational
interactions compared to elementary particle interactions, this is
expected to be a reasonable approximation which nevertheless has a
range of applicability where nontrivial phenomena occur, like particle
creation in strong, or rapidly varying, gravitational fields. The most
prominent effects of that sort are the Hawking effect \cite{Haw.eff} and the
Fulling-Unruh effect \cite{Ful.eff,Unr.eff}. 

For quantum field theory on Minkowski spacetime, one demands that
quantum fields behave covariantly under Poincar\'e-transformations,
and there are distinguished states, like the vacuum state, or (multi-)
particle states tied to the Wigner-type particle concept. Such states
are natural reference states which
allow it to fix physical quantities in comparison with experiments. In
contradistinction to this familiar case, a generic spacetime manifold
need not possess any (non-trivial) spacetime symmetries (isometries),
and thus there is in general no restrictive concept of covariance for
quantum fields propagating on an arbitrary, but fixed curved
background spacetime. (A similar problem arises already for quantum
fields in flat spacetime coupled to outer classical fields, and 
most of what follows applies, mutatis mutandis, also to this case.)

This lack of covariance is a source of serious ambiguities in quantum
field theory on curved spacetime, such as the lack of a natural
candidate of a vacuum state or a Wigner-type particle concept. In
turn, this leads to ambiguities in the concrete determination of physical
quantities. This problem was observed some time ago
by Wald \cite{Wald2} in his discussion of a
renormalization prescription for defining the energy-momentum tensor
of a quantized field on a curved spacetime $M$ with metric tensor $\gb
= g_{\mu\nu}$. 
For a classical massless Klein-Gordon 
field $\varphi$, the canonical energy-momentum tensor is at $x \in M$
\begin{equation*}
T_{\mu\nu}(x) =  \nabla_\mu \varphi(x) \nabla_\nu
\varphi(x)
 -\frac{1}{2}g_{\mu\nu}(x)
\nabla_\lambda\varphi(x)\nabla^\lambda\varphi(x) \,,
\end{equation*}
where $\nabla$ denotes the metric-covariant derivative. 
For a quantum field $\varphi$ pointwise products are ill-defined, and therefore
one starts from the point-split expression 
\begin{equation*}
T_{\mu\nu}(x,y) = \left(\nabla_\mu \varphi(x) \nabla_\nu
  \varphi(y)-  \frac{1}{2}
g_{\mu\nu}(x,y)(g^{\rho\sigma}(x,y)\nabla_\rho\varphi(x)  
 \nabla_\sigma\varphi(y))\right)   
\end{equation*}
where $g_{\mu\nu}(x,y)$ is the bitensor obtained from $g_{\mu\nu}$ by parallel
transport along the geodesic from $x$ to $y$. Since we are eventually 
interested in the coincidence limit $y \to x$ we may consider the
points as near as  
needed to get a unique geodesic, hence we do not have any problem of choices. 

The idea is now to subtract from $T_{\mu\nu}(x,y)$ a suitable scalar 
distribution $t_{\mu\nu}(x,y)$ such that the renormalized
energy-momentum tensor 
\begin{equation*}
T_{\mu\nu}^{\rm ren}(x) =\lim_{y \to x}\, \left(T_{\mu\nu}(x,y)-
  t_{\mu\nu}(x,y)\right) 
\end{equation*}
may be defined by a well-determined coincidence limit $y \to x$.
Inspired by the similar situation in Minkowski spacetime,
a first approach of going about this may be taken as follows:
Choose a quasi-free Hadamard state $\omega$ as ``reference state'', and let
\begin{equation*}
t_{\mu\nu}(x,y) = \omega(T_{\mu\nu}(x,y)) \ .
\end{equation*}
Then $T_{\mu\nu}^{\rm ren}(x)$ exists as a well defined operator 
valued distribution in all representations induced by arbitrary
(other) Hadamard states \cite{BFK}. 
The problem
of this definition is the non-uniqueness of the chosen reference
state: on a generic curved spacetime, there are plenty of quasifree
Hadamard states, and none of them is distinguished in the way the
vacuum state is on Minkowski spacetime owing to the circumstance that a
generic curved spacetime need not admit non-trivial (conformal)
isometries. Therefore, choosing different reference states,
one gets quite arbitrary modifications of $T_{\mu\nu}^{\rm ren}(x)$ in the
form of added 
symmetric numerical tensors; hence, little can be said in this 
approach, for instance, about the back reaction of quantum matter on
the gravitational field.

In order to restrict this ambiguity, Wald imposed as a further
requirement a principle of locality and covariance 
which states that the energy-momentum tensor should only locally depend on the
spacetime metric; we will outline this condition further below.
 Starting from this principle he gave a definition of the
subtraction term $t_{\mu\nu}$ which depends only on the local metric. By this
method he found a covariantly conserved energy momentum tensor, and, as 
a byproduct,
the conformal anomaly showed up, namely, in the case of the conformally
covariant Klein-Gordon field it is not possible to find a 
$t_{\mu\nu}$ such that the resulting energy-momentum tensor is both
conserved and traceless. The ambiguity in the definition of the
renormalized energy-momentum tensor is now reduced to a local curvature
term \cite{Wald2}.  

A similar problem occurred in the definition of Wick-polynomials and of 
renormalized perturbation theory on Lorentzian manifolds. For instance, the 
definition of the Wick square $:\!\varphi^2\! :_\omega \! (x)$ given
in \cite{BFK} also takes the form of a coincidence limit
\begin{equation*}
:\!\varphi^2\! :_{\omega}\! (x)=\lim_{y\to x} (\varphi(x)\varphi(y)-
\omega(\varphi(x)\varphi(y)))\ ,
\end{equation*}
where $\omega$ is some fixed reference
state (again being a quasifree Hadamard state; the limit 
procedure has to be properly defined, see, e.g.\ \cite{BFK}). 
Again due to the non-unique choice of a reference state, it turns out
that chosing instead of $\omega$ a different reference state $\omega'$
results in 
$:\!\varphi^2\! :_{\omega}\! (x)$ to $:\!\varphi^2\! :_{\omega}\! (x)+ f(x)$
with some smooth function $f$.
This ambiguity would actually not be very serious at the level of a
description of a quantum field theory in terms of operator algebras,
but it enters into the definition of time-ordered products of
Wick-polynomials from which, in turn, local $S$-matrix functionals are
derived in the sense of perturbation theory whose matrix elements may
be compared with physical processes modelled by interacting fields on
curved spacetime \cite{BF}. Furthermore, a more serious ambiguity
enters in the course of the process of infinite renormalization of
ultraviolet divergencies in defining the time-ordered product of
Wick-polynomials. There remains a freedom that corresponds to adding
certain products of differential operators contracted with Wick-polynomials to
the Lagrangean. While one can show \cite{BF} that the perturbative
classification of interacting scalar field theories on curved
spacetimes is independent of that freedom, the predictive power of the
local $S$-matrix thus obtained is somewhat limited because the
``renormalization constants'' now are, in fact, functions depending on
the spacetime points. Therefore, it seems most desirable to invoke a
suitable locality and covariance principle so as to reduce that
ambiguity affecting the $S$-matrix in a similar way as was done by
Wald for the case of the energy-momentum tensor. And, in fact, in
recent work by Hollands and Wald \cite{HW}, this task has been
attacked successfully. We should like to point out that related ideas
concerning the renormalization of physical quantities for quantum
fields in flat spacetime coupled to outer electromagnetic fields have
been proposed earlier by Dosch and M\"uller \cite{DM}.

Let us now briefly look at the locality and 
covariance condition imposed by Wald
\cite{Wald2} in order to reduce the ambiguity of the renormalized
energy-momentum tensor of the free, massless scalar field. The
condition may be formulated as follows. Suppose that one has a
prescription for obtaining $T^{\rm ren}_{\mu \nu}(x)$ on {\it any}
curved spacetime. Then such a prescription is local and covariant if
the following holds: Whenever one has two spacetimes $M$ and $M'$
equipped with metrics $\gb$ and $\gb'$, respectively, and for some
(arbitrary) open subset $U$ of $M$ an isometric diffeomorphism $\kappa
: U \to U'$ onto an open subset $U'$ of $M'$ (so that $\kappa_*\gb =
\gb'$), then it is required that 
$$ \alpha'_{\kappa}(T'{}^{\rm ren}_{\mu\nu}(x')) = \kappa_*T^{\rm
  ren}_{\,\mu\nu} (x')\,, \quad x' \in U'\,,$$
where $\alpha'_{\kappa}: \calA_{M'}(U') \to \calA_{M}(U)$ is the
canonical isomorphism between the local CCR-algebras $\calA_{M'}(U')$
of the Klein-Gordon field on $M'$ and $\calA_{M}(U)$ of the
Klein-Gordon field on $M$ (cf.\ \cite{Dim.KG,Wald2}), and
$T_{\mu\nu}^{\rm ren}$ is the renormalized energy momentum tensor
according to the renormalization prescription on $M$, and $T'{}^{\rm
  ren}_{\mu\nu}$ that on $M'$ according to the same prescription. In
other words, the condition demands that $\kappa_*T^{\rm ren}_{\mu\nu}
= T'{}^{\rm ren}_{\mu \nu}$ up to a canonical algebraic isomorphism
(strictly speaking, this is only valid at the level of expectation
values in Hadamard states).

Two things should be noted. First, the neighbourhood $U$ was
arbitrary, and therefore the information entering into the above
condition is local (in the sense of being independent of what happens
in the surroundings of $U$ or $U'$). Secondly, the condition makes
considerable use of the fact that the quantum theory of the free
scalar field can be formulated on every (globally hyperbolic) spacetime
and that there is a canonical way of identifying the corresponding
quantum field theories on isometrically diffeomorphic subregions of
globally different 
spacetimes by an algebraic isomorphism $\alpha'_{\kappa}$. Quantum
field theories of that kind respect the dictum of general relativity
to regard events (quantum processes) simultaneously as taking place in
several, locally isomorphic spacetimes. 

The further formalization of
this property is the main purpose of the present article.
The most general and most efficient mathematical framework for such a
discussion is provided by the operator-algebraic approach to quantum
field theory which was initiated by Haag and Kastler \cite{HK} for
quantum field theory on Minkowski spacetime, see also the monographs
\cite{Haag,Araki.book}. 
In Section \ref{gclp}, we will define a local, generally
covariant quantum field theory as a covariant functor between the category of
globally hyperbolic (four-dimensional) spacetime manifolds with
isometric embeddings as morphisms and the category of $C^*$-algebras
with invertible endomorphisms as morphisms. This generalizes similar
approaches, such as the notion of a local, covariant quantum field
recently used in \cite{HW}, and is very similar to the concept of
a covariant field
theory over the class of globally hyperbolic manifolds defined in
\cite{Verch2}.  The latter is a generalization of ideas in
\cite{Dim.D} where also the setting of categories and  functors was
used. Our approach seems to have the advantage of generalizing in a
natural manner  at the
same time all these mentioned concepts as well as related ideas on
generally covariant quantum field theories which appear e.g.\ in the
famous ``Missed opportunities'' collection by Dyson \cite{Dyson}, or
in the works \cite{Ban,FH,Haag}.
We will indicate that the theory of a free, scalar
Klein-Gordon field on globally hyperbolic spacetimes is an example for
our functorial description of a quantum field theory.
 Moreover, it will turn out that the
more common concept of a quantum field theory on a fixed spacetime
background described in terms of an isotonous map from bounded open
subregions to $C^*$-algebras which is covariant when the spacetime
possesses isometries (as in the original Haag-Kastler approach on
Minkowski-spacetime, as will be indicated below) 
is actually a consequence of our functorial description.

Section 3 is devoted to a study of the functorial properties of
state spaces for locally covariant quantum field theories. A state
space will be introduced as a contravariant functor between the
category of globally hyperbolic spacetimes and the category of dual
spaces of $C^*$-algebras, with duals of $C^*$-algebraic embeddings as
morphisms. State spaces will be characterized which have the property
that their ``local folia'' are left invariant under the functorial
action of isometric embeddings of spacetime manifolds. These will be
seen to obey the principle of local definiteness proposed by Haag,
Narnhofer and Stein \cite{HNS}. We will indicate that the quasifree
states of the Klein-Gordon field which fulfill the microlocal spectrum
condition \cite{BFK} or equivalently, the Hadamard condition
\cite{Rad1,KW}, induce such a state space.

In Section 4 we will demonstrate that for locally covariant quantum
field theories obeying the time-slice axiom one can associate a
dynamics in the form of automorphic actions, referred to as ``relative
Cauchy-evolution'', which describe the reaction of the quantum field
theory on local perturbations of the spacetime metric. We will show
that the functional derivative of such relative Cauchy-evolutions with
respect to the spacetime-metric is divergence-free. This functional
derivative has, in analogy to the case of classical field theory, the
significance of an energy-momentum tensor, and in fact we will also
show that for the free Klein-Gordon field the functional derivative of
the relative Cauchy-evolution agrees with the commutator action of the
energy momentum tensor in representations of quasifree Hadamard states.  

Finally, in Section 5, as
an alternative to the approach by Hollands and Wald \cite{HW}, we will
point out that the construction of local, covariant Wick-polynomials
of the free field arises as solution of a simple cohomological problem. 

Some technical details appear in an Appendix.

\section{The Generally Covariant Locality Principle}\label{gclp}
\subsection{Some Geometrical Preliminaries}
${}$\\[6pt]
In the following, we will be concerned with four-dimensional, globally
hyperbolic spacetimes, so it is in order to summarize some of their
basic properties. For further discussion, see e.g.\ \cite{HE,Wald1}.
We note that the condition of global hyperbolicity doesn't appear to
be very restrictive on physical grounds. Its main purpose is to rule
out certain causal pathologies.

We denote a spacetime by $(M,\gb)$ where $M$ is a smooth,
four-dimensional manifold (smooth meaning here $C^{\infty}$, and
Hausdorff, paracompact, and connected) and $\gb$ is a Lorentzian metric on $M$
(taken to be of signature $(+1,-1,-1,-1)$). Also, we always assume
that the spacetimes we consider are orientable and
time-orientable. The latter means that there exists a
$C^{\infty}$-vectorfield $u$ on $M$ which is everywhere timelike,
i.e.\ $\gb(u,u) > 0$. A smooth curve $\gamma: I \to M$, $I$ being a
connected subset of $\R$, is called causal if
$\gb(\dot{\gamma},\dot{\gamma}) \ge 0$ where $\dot{\gamma}$ denotes
the tangent vector of $\gamma$. Given the global timelike vectorfield
$u$ on $M$, one calls a causal curve $\gamma$ future-directed if
$\gb(u,\dot{\gamma}) > 0$ all along $\gamma$, and analogously one
calls $\gamma$ past-directed if $\gb(u,\dot{\gamma}) < 0$. This
induces a globally consistent notion of time-direction in the spacetime
$(M,\gb)$. For any point $x \in M$, $J^{\pm}(x)$ denotes the set of
all points in $M$ which can be connected to $x$ by a
future(+)/past$(-)$-directed causal curve $\gamma: I \to M$ so that $x
= \gamma(\inf \,I)$. Two subsets $O_1$ and $O_2$ in $M$ are called
causally separated if they cannot be connected by a causal curve, i.e.\
if for all $x \in \overline{O_1}$,
 $J^{\pm}(x)$ has empty intersection with $\overline{O_2}$.
By $O^{\perp}$ we denote the causal complement of $O$, i.e.\ the
largest open set in $M$ which is causally separated from $O$.

An orientable and time-orientable spacetime $(M,\gb)$ is called
globally hyperbolic if for each pair of points $x,y \in M$ the set
$J^-(y) \cap J^+(x)$ is compact whenever it is non-empty. This
property can be shown to be equivalent to the existence of a smooth
foliation of $M$ in Cauchy-surfaces, where a smooth hypersurface of
$M$ is called a Cauchy-surface if it is intersected exactly once by
each inextendible causal curve in $(M,\gb)$ (for precise definition of
inextendible causal curve, see the indicated references).
A particular feature of globally hyperbolic spacetimes is the
fact that the Cauchy-problem (inital value problem) for linear
hyperbolic wave-equations is well-posed and that such 
wave-equations possess unique retarded and advanced fundamental
solutions on those spacetimes. It should also be observed that global
hyperbolicity makes no reference to spacetime isometries.

Of some importance later on will be the concept of isometric
embedding. Let $(M_1,\gb_1)$ and $(M_2,\gb_2)$ be two globally
hyperbolic spacetimes. A map $\psi: M_1 \to M_2$ is called an isometric
embedding (of $(M_1,\gb_1)$ into $(M_2,\gb_2)$) if $\psi$ is a
diffeomorphism onto its range $\psi(M_1)$ (i.e.\ the map $\bar{\psi}:
M_1 \to \psi(M_1) \subset M_2$ is a diffeomorphism) and if $\psi$ is
an isometry, that is, $\psi_*\gb_1 = \gb_2 \rest\psi(M_1)$.

\subsection{Quantum Field Theories as Covariant Functors}
${}$\\[6pt]
It is a famous saying attributed to E.\ Nelson that quantum field theory is a 
functor. This has to do with the map of second 
quantization, mapping the category of Hilbert-spaces 
with unitaries as morphisms to that of C$^\ast$-algebras with
unit-preserving $\ast$-homomorphisms as morphisms. 
In a similar light, topological quantum field theories have already at an
early stage been couched in the framework of categories and functors
\cite{Atiyah}. Here, we wish to put forward that quantum field theory
is indeed a covariant functor, but in the more fundamental and physical
sense of implementing the principles of locality and general
covariance, as discussed in the Introduction. As already pointed out, 
our approach provides a natural generalization
both of the usual abstract formulation of quantum field theory in
terms of isotonous families of operator algebras indexed by bounded
open subregions of a fixed background spacetime, and of other
approaches to diffeomorphism-covariant quantum field theory; we will
discuss this further below. We first have to define the categories
involved in our formulation of locally covariant quantum field
theory. (See \cite{MacLane} as general reference on categories and functors.)
The two categories we shall use are the following:
\begin{description}
\item[$\frakM$] This category consists of a class of objects
$\obj(\frakM)$ formed by all four-dimensional, globally
hyperbolic spacetimes $(M,\gb)$ which are oriented and time-oriented.
Given any two such objects $(M_1,\gb_1)$ and $(M_2,\gb_2)$, the morphisms
$\psi\in\Hom_{\frakM}((M_1,\gb_1),(M_2,\gb_2))$ 
are taken to be the isometric embeddings
$\psi: (M_1,\gb_1) \to (M_2,\gb_2)$ of $(M_1,\gb_1)$ into
$(M_2,\gb_2)$
 as defined above, but with the additional 
constraints that 
\begin{itemize}
\item[$(i)$] if $\gamma : [a,b]\to M_2$ is any 
causal curve and
$\gamma(a),\gamma(b)\in\psi(M_1)$ then the whole curve must be 
in the image $\psi(M_1)$, i.e., $\gamma(t)\in\psi(M_1)$ for all $t\in ]a,b[$; 
\item[$(ii)$] the isometric embedding preserves orientation and
  time-orientation of the embedded spacetime.
\end{itemize} 
The composition rule for any $\psi \in
\Hom_{\frakM}((M_1,\gb_1),(M_2,\gb_2))$ and $\psi' \in$
\linebreak
 $\Hom_{\frakM}((M_2,\gb_2),(M_3,\gb_3))$ is to define its composition
$\psi' \circ \psi$ as the composition of maps. Hence
$\psi' \circ \psi: (M_1,\gb_1) \to (M_3,\gb_3)$ is a well-defined map
which is obviously a diffeomorphism onto its range $\psi'(\psi(M_1))$
and clearly isometric; also the properties $(i)$ and and $(ii)$ are
obviously fulfilled, and hence $\psi' \circ \psi \in
\Hom_{\frakM}((M_1,\gb_1),(M_3,\gb_3))$. The associativity of the
composition rule follows from the associativity of the composition of
maps. Clearly, each $\Hom_{\frakM}((M,\gb), (M,\gb))$ possesses a unit
element, given by the identical map ${\rm id}_M : x \mapsto x$, $x \in
M$. 
\end{description}

\begin{description}
\item[$\frakA$] This is the category whose class of objects
  $\obj(\frakA)$ is formed by all $C^*$-algebras possessing unit
  elements, and the morphisms are faithful (injective) unit-preserving
  $*$-homomorphisms. Given $\alpha \in \Hom_{\frakA}(\calA_1,\calA_2)$
  and $\alpha' \in \Hom_{\frakA}(\calA_2,\calA_3)$, the composition
  $\alpha' \circ \alpha$ is again defined as the composition of maps
  and easily seen to be an element in
  $\Hom_{\frakA}(\calA_1,\calA_3)$. The unit element in
  $\Hom_{\frakA}(\calA,\calA)$ is for any $A \in \obj(\frakA)$ given
  by the identical map ${\rm id}_{\calA}: A \mapsto A$, $A\in\calA$.
\end{description}
{\it Remarks. } (A) Requirement $(i)$ on the morphisms of $\frakM$ is
introduced in order that the induced and intrinsic causal structures
coincide for the embedded spacetime $\psi(M_1) \subset M_2$. Aspects
of this condition are discussed in \cite{Kay.RomePr}. Condition
$(ii)$ might, in fact, be relaxed; the resulting structure, allowing
also isometric embeddings which reverse spatial- and time-orientation,
could accomodate a discussion of PCT-theorems. We hope to report
elsewhere on this topic.
\\[6pt]
(B) Clearly, one may envisage variations on the categories introduced
here. Our present choices might have to be changed or supplemented by
other structures, depending on the situations considered. For example,
instead of choosing for $\obj(\frakA)$ the class of $C^*$-algebras
with unit elements, one could consider $*$-algebras,
Borchers-algebras, or von Neumann algebras; we have chosen
$C^*$-algebras for definiteness. Moreover, one could also allow more
general objects than globally hyperbolic spacetimes in $\obj(\frakM)$,
or endow these objects with additional structures, e.g.\
spin-structures, as in \cite{Dim.D,Verch2}. For discussing the
locality and covariance structures of observables, however, the present
approach appears sufficient.
\\[6pt]
Now we wish to define the concept of locally covariant quantum field theory.
\begin{Def} \label{CovFunctor}
${}$\\
$(i)$\quad
A {\bf locally covariant quantum field theory} is a
covariant functor $\euA$ between the two categories $\frakM$ and $\frakA$, 
i.e., writing $\alpha_{\psi}$ for $\euA(\psi)$, in typical diagramatic form:
\begin{equation*}
\begin{CD}
(M,\gb) @>\psi>> (M',\gb')\\
@V{\euA}VV     @VV{\euA}V\\
\euA(M,\gb)@>{\alpha_\psi}>> \euA(M',\gb')
\end{CD}
\end{equation*}
together with the covariance properties
$$ \alpha_{\psi'} \circ \alpha_{\psi} = \alpha_{\psi' \circ \psi}\,,
\quad \alpha_{{\rm id}_M} = {\rm id}_{\euA(M,\gb)}\,,$$
for all morphisms 
$\psi \in \Hom_{\frakM}((M_1,\gb_1),(M_2,\gb_2))$, all morphisms $\psi' \in
$ \linebreak
 $\Hom_{\frakM}((M_2,\gb_2),(M_3,\gb_3))$ and all $(M,\gb) \in
 \obj(\frakM)$.
\\[6pt]
$(ii)$\quad A locally covariant quantum field
theory described by a covariant functor $\euA$ is called {\bf causal}
if the following holds: Whenever there 
are morphisms $\psi_j \in \Hom_{\frakM}((M_j,\gb_j),(M,\gb))$, $j
=1,2$, so that the sets $\psi_1(M_1)$ and $\psi_2(M_2)$ are causally
separated in $(M,\gb)$, then one has
$$ \left[
  \alpha_{\psi_1}(\euA(M_1,\gb_1)),\alpha_{\psi_2}(\euA(M_2,\gb_2))\right]
 = \{0\}\,, $$
where $[\calA,\mathcal{B}] = \{AB-BA: A \in \calA, B \in
\mathcal{B}\}$ for any pair of $C^*$-algebras $\calA$ and
$\mathcal{B}$.
\\[6pt]
$(iii)$\quad 
We say that a locally covariant quantum
field theory given by the functor $\euA$ obeys the {\bf time-slice axiom} if
$$ \alpha_{\psi}(\euA(M,\gb)) = \euA(M',\gb') $$
holds for all $\psi \in \Hom_{\frakM}((M,\gb),(M',\gb'))$ such that
$\psi(M)$ contains a Cauchy-surface for $(M',\gb')$.
\end{Def}
Thus, a  locally covariant
quantum field theory is an assignment of $C^*$-algebras to
(all) globally hyperbolic spacetimes so that the algebras are
identifyable when the spacetimes are isometric, in the indicated way.
Note that we use the term ``local'' in the sense of  ``geometrically
local" in the definition which shouldn't be confused
 with the meaning of locality in the sense of Einstein causality.
Causality properties are further specified in $(ii)$ and $(iii)$ of 
Def.\ \ref{CovFunctor}. Causality means that the algebras
$\alpha_{\psi_1}(\euA(M_1,\gb_1))$  and
$\alpha_{\psi_2}(\euA(M_2,\gb_2))$ commute elementwise in the larger
algebra $\euA(M,\gb)$ when the sub-regions $\psi_1(M_1)$ and
$\psi_2(M_2)$ of $M$ are causally separated (with respect to
$\gb$). This property is expected to hold generally for observable
quantities which can be localized in certain subregions of
spacetimes. The time slice axiom $(iii)$ (also called strong Einstein
causality, or existence of a causal dynamical law, cf.\ \cite{Verch2})
says that an algebra of observables on a globally hyperbolic
spacetime is already determined by the algebra of observables localized
in any neighbourhood of a Cauchy-surface.

 Before continuing, some remarks on related
approaches are in order now. In \cite{Dyson}, Dyson suggested that
one should attempt to generalize the usual Haag-Kastler framework of a
general description of quantum field theories on Minkowski spacetime,
as we have sketched it in the Introduction, to general spacetime
manifolds in such a way that the covariance group $\Poif$ is replaced
by the diffeomorphism group. An approach which is very close in spirit
to Dyson's suggestion is due to Bannier \cite{Ban} who constructed, on
$\R^4$ as fixed background manifold, a generalized CCR-algebra of the
Klein-Gordon field of fixed mass on which the diffeomorphism group
acts covariantly by $C^*$-automorphisms. Bannier's approach may
therefore be regarded as a realization of a functor $\euA$ with the
above properties but where the domain-category $\frakM$ is replaced by
the subcategory $\frakM_{\R^4}$ whose objects are the globally
hyperbolic spacetimes $(M,\gb)$ having $M = \R^4$ as spacetime
manifolds, and globally hyperbolic sub-spacetimes of those. However, it
appears that the restriction to a fixed background manifold like
$\R^4$ is artificial, and at variance with the principles of general
relativity. This is supported by the results in \cite{Verch2} where an
approach similar to the one presented here was taken, and which
``localizes'' Dimock's formulation in \cite{Dim.KG,Dim.D} where a functorial
approach to generally covariant quantum field theory seems to have
been proposed for the first time. Like Bannier's work, however,
Dimock's proposal lacks the ``locality'' aspect of general covariance
and therefore doesn't completely reveal its strength. It was shown in
\cite{Verch2} that the combination of general covariance and
(geometrical) locality leads, together with a few other, natural
requirements, to a spin-statistics theorem for quantum fields on
curved spacetimes. 
\subsection{The Klein-Gordon Field}\label{kgfe}
${}$\\[6pt]
The simplest and best studied example of a quantum field theory in
curved spacetime is the scalar Klein-Gordon field. As was shown by
Dimock \cite{Dim.KG}, its local $C^*$-algebras can be constructed
easily on each globally hyperbolic spacetime, giving rise to a functor
$\euA$. To summarize this construction, let $(M,\gb)$ be an object in
$\obj(\frakM)$. Global hyperbolicity entails the well-posedness of the
Cauchy-problem for the scalar Klein-Gordon equation on $(M,\gb)$,
\begin{equation}\label{KGeqn}
(\nabla^a\nabla_a + m^2 + \xi R)\varphi = 0
\end{equation}
(for smooth, real-valued $\varphi$) where $\nabla$ is the covariant
derivative of $\gb$, $m \ge 0$ and $\xi \ge 0$ are constants, and $R$
is the scalar curvature of $\gb$. Moreover, it implies that there
exist uniquely determined advanced and retarded fundamental solutions
of the Klein-Gordon equation, $E^{\rm adv/ret}: C_0^{\infty}(M,\R) \to
C^{\infty}(M,\R)$. Their difference $E= E^{\rm adv} -E^{\rm ret}$ 
is called the causal
propagator of the Klein-Gordon equation. Let us denote the range
$E(C^{\infty}_0(M,\R))$ by $\mathcal{R}$ (or, sometimes, by
$\mathcal{R}(M,\gb)$ for clarity). It can
be shown (cf.\ \cite{Dim.KG}) that defining
$$ \sigma(Ef,Eh) = \int_M f(Eh)\,d\mu_{\gb}\,, \quad f,h \in
C^{\infty}_0(M,\R)\,,$$
where $d\mu_{\gb}$ is the metric-induced volume form on $M$, endowes
$\mathcal{R}$ with a symplectic form, and thus $({\mathcal R},\sigma)$ is
a symplectic space. To this symplectic space one can associate its
Weyl-algebra $\mathfrak{W}({\mathcal R},\sigma)$, which is generated
by a family of unitary elements $W(\varphi)$, $\varphi \in {\mathcal R}$,
satisfying the CCR in exponentiated form (``Weyl-relations''),
$$ W(\varphi)W(\tilde{\varphi}) = {\rm
  e}^{-i\sigma(\varphi,\tilde{\varphi})/2} W(\varphi + \tilde{\varphi})\,.$$
Now, when the constants $m$ and $\xi$ are kept fixed independently of
$(M,\gb)$, the symplectic space $({\mathcal R},\sigma)$ is 
entirely determined by
$(M,\gb)$, and so is $\mathfrak{W}({\mathcal R},\sigma)$. Setting
therefore $\euA(M,\gb) = \mathfrak{W}({\mathcal
  R}(M,\gb),\sigma_{(M,\gb)})$,
 one obtains a candidate for a 
covariant functor $\euA$ with the properties of Def.\ 2.1.
What remains to be checked is the covariance property. Thus, let $\psi
\in \Hom_{\frakM}((M,\gb),(M',\gb'))$ and let us denote by
$E,{\mathcal R},\sigma$ the propagator, range-space, and symplectic
form corresponding to the Klein-Gordon equation \eqref{KGeqn}
on $(M,\gb)$, and by $E',{\mathcal R}',\sigma'$ their counterparts
with respect to $(M',\gb')$. Moreover, we denote by
$E^{\psi},{\mathcal R}^{\psi},\sigma^{\psi}$ the analogous objects for
the spacetime $(\psi(M),\psi_*\gb)$. It was shown in \cite{Dim.KG}
that, writing $\psi_*\varphi = \varphi \circ \psi^{-1}$, there holds
$E^{\psi} = \psi_* \circ E \circ \psi_*{}^{-1}$, ${\mathcal R}^{\psi}
= \psi_*{\mathcal R}$, and $\sigma(Ef,Eh) =
\sigma^{\psi}(E^{\psi}\psi_*f,E^{\psi}\psi_*h) =
\sigma^{\psi}(\psi_*Ef,\psi_*Eh)$. Thus $\psi_*$ furnishes a
symplectomorphism between $({\mathcal R},\sigma)$ and $({\mathcal
  R}^{\psi},\sigma^{\psi})$, and hence, by a standard theorem
\cite{BR1}, there is a $C^*$-algebraic isomorphism
$\tilde{\alpha}_{\psi}: \mathfrak{W}({\mathcal R},\sigma) \to
\mathfrak{W}({\mathcal R}^{\psi},\sigma^{\psi})$ so that
\begin{equation}
\label{Bog1}
 \tilde{\alpha}_{\psi}(W(\varphi)) = W^{\psi}(\psi_*(\varphi))\,, \quad
 \varphi \in {\mathcal R}\,
\end{equation}
where $W^{\psi}(\,.\,)$ denote the CCR-generators of
$\mathfrak{W}({\mathcal R}^{\psi},\sigma^{\psi})$. 

While these observations are already contained in Dimock's work
\cite{Dim.KG}, we add another one which is important in the present
context: Since $\psi: M \to \psi(M) \subset M'$ is a metric isometry,
it holds that $\psi_*\gb = \gb'\rest \psi(M)$. And hence the fact that
the advanced and retarded fundamental solutions of the Klein-Gordon
operator are uniquely determined on a globally hyperbolic spacetime
implies that $E^{\psi} =\chi_{\psi(M)} E'\rest C^{\infty}_{0}(\psi(M),\R)$
where $\chi_{\psi(M)}$ is the characteristic function of $\psi(M)$ and
that, moreover, ${\mathcal R}^{\psi}$ can be identified with 
$E'(C_0^{\infty}(\psi(M),\R))$ and $\sigma^{\psi}$ with
$\sigma'\rest {\mathcal R}^{\psi}$. 
Therefore, the map $T^{\psi}$ which assigns to each element $Ef$, $f
\in C_0^{\infty}(M,\R)$, the element $E'\iota_{\psi}{}_*f$ in
$(\mathcal{R}',\sigma')$, is a symplectic map from
$(\mathcal{R}^{\psi},\sigma^{\psi})$ into $(\mathcal{R}',\sigma')$,
and thus one
obtains a $C^*$-algebraic endomorphism $\tilde{\alpha}_{\iota_{\psi}}:
\mathfrak{W}({\mathcal R}^{\psi},\sigma^{\psi}) \to
\mathfrak{W}({\mathcal R}',\sigma')$ by
\begin{equation} \label{Bog2}
\tilde{\alpha}_{\iota_{\psi}}(W^{\psi}(\phi)) =
W'(T^{\psi}\phi)\,,  \quad \phi \in {\mathcal R}^{\psi}\,,
\end{equation}
where $W'(\,.\,)$ denote the Weyl-generators of
$\mathfrak{W}({\mathcal R}',\sigma')$. Hence, setting
$\alpha_{\psi} =
\tilde{\alpha}_{\iota_{\psi}}\circ\tilde{\alpha}_{\psi}$, we have a
$C^*$-algebraic endomorphism $\alpha_{\psi}: \euA(M,\gb) \to
\euA(M',\gb')$. The covariance property $\alpha_{\psi' \circ \psi} =
\alpha_{\psi'} \circ \alpha_{\psi}$ for $\psi \in
\Hom_{\frakM}((M,\gb),(M',\gb'))$ and $\psi' \in
\Hom_{\frakM}((M',\gb'),(M'',\gb''))$ is an easy consequence of the
construction of $\alpha_{\psi}$, i.e.\ of the relations \eqref{Bog1} and
\eqref{Bog2}. 
It was also shown in \cite{Dim.KG} that causality and time-slice axiom
are fulfilled in each $\mathfrak{W}(\mathcal{R},\sigma)$ in the
following sense: (i) If $f,h \in C^{\infty}_0(M,\R)$ with ${\rm
  supp}\,f \subset ({\rm supp}\,h)^{\perp}$, then $W(Ef)$ and $W(Eh)$
commute, (ii) if $N$ is an open neighbourhood of a Cauchy-surface
$\Sigma$ in $M$, then there is for each $f\in C^{\infty}_0(M,\R)$
some $h \in C^{\infty}_0(N,\R)$ with $W(Ef) = W(Eh)$.
We collect these findings in the following
\begin{Thm}
If one defines for each $(M,\gb) \in \obj(\frakM)$ the $C^*$-algebra
$\euA(M,\gb)$ as the CCR-algebra $\mathfrak{W}({\mathcal
  R}(M,\gb),\sigma_{(M,\gb)})$ of the Klein-Gordon equation
\eqref{KGeqn} (with $m,\xi$ fixed for all $(M,\gb)$), and for each
$\psi \in \Hom_{\frakM}((M,\gb),(M',\gb'))$ the $C^*$-algebraic
endomorphism $\alpha_{\psi} = \tilde{\alpha}_{\iota_{\psi}} \circ
\tilde{\alpha}_{\psi}: \euA(M,\gb) \to \euA(M',\gb')$ according to
\eqref{Bog1} and \eqref{Bog2}, then one obtains in this way a
covariant functor $\euA$ with the properties of Def.\
\ref{CovFunctor}.
Moreover, this functor is causal and fulfills the time-slice axiom.
\end{Thm}
In this sense, the free Klein-Gordon field theory is a locally covariant
 quantum field theory.
\subsection{Recovering Algebraic Quantum Field Theory}\label{raqft}
${}$\\[6pt]
Now, we will address the issue of re-gaining the usual
setting of algebraic quantum field theory on a fixed globally
hyperbolic spacetime from a locally covariant quantum field theory, 
i.e.\ from a covariant functor $\euA$ with
the properties listed above. 
It may be helpful for readers not too familiar with the algebraic
approach to quantum field theory on Minkowski spacetime that we
briefly summarize the Haag-Kastler framework \cite{HK} so that it
becomes apparent in which way the usual description of algebraic
quantum field theory is re-gained via Prop.\ \ref{localnet} from
our functorial approach. 
In the Haag-Kastler framework, the basic structure of
the formal description of a quantum system is given by a map $O
\mapsto {\calA}(O)$ assigning to each open, bounded region $O$ a
$C^*$-algebra ${\calA}(O)$. This ``local $C^*$-algebra'' is supposed
to contain all the (bounded) observables of the quantum system at hand
that can be measured ``at times and locations'' within the spacetime
region $O$; e.g., if the system is described by a hermitean scalar 
quantum field
$\varphi(x)$, then ${\calA}(O)$ may be taken as the operator-algebra
generated by all exponentiated field operators ${\rm e}^{i\varphi(f)}$
where the test-functions $f$ are supported in $O$, and the smeared
field-operators are $\varphi(f) = \int d^4x\,f(x)\varphi(x)$. Hence,
one has the condition of isotony, demanding that
${\calA}(O_1) \subset {\calA}(O)$ whenever $O_1 \subset
O$. It is also assumed that the local algebras all contain a common
unit element, denoted by ${\bf 1}$. 
Moreover, as the local algebras contain observables, it is usually
demanded that they commute elementwise when their respective
localization regions are spacelike separated. 

The locality concept
being thus formulated, the notion of special relativistic covariance
is given the following form: Collecting all local observables in the
minimal $C^*$-algebra ${\calA}$ containing all local algebras 
${\calA}(O)$,\,\footnote{This minimal $C^*$-algebra is, as a consequence of
  the isotony condition, well-defined and in the mathematical
  terminology called the inductive limit of the family $\{{\calA}(O)\}$
  where $O$ ranges over all bounded open subsets of Minkowski
  spacetime.}
there ought to be for each element $L \in \Poif$ (i.e., the proper,
orthochronous Poincar\'e group) a $C^*$-algebra automorphism $\alpha_L
: {\calA} \to {\calA }$ so that
$$ \alpha_{L_1} \circ \alpha_{L_2} = \alpha_{L_1 \circ L_2}\,, \quad
L_1,L_2 \in \Poif\,,$$
where $L_1 \circ L_2$ denotes the composition of elements in
$\Poif$. 

Let $(M,\gb)$ be an object
in $\obj(\frakM)$. We denote by $\calK(M,\gb)$ the set of all subsets
in $M$ which are relatively compact and contain with each pair of
points $x$ and $y$ also all $\gb$-causal curves in $M$ connecting $x$
and $y$ (cf.\ condition $(ii)$ in the definition of $\frakM$). Given
$O \in \calK(M,\gb)$, we denote by $\gb_O$ the Lorentzian metric
restricted to $O$, so that $(O,\gb_O)$ (with the induced orientation
and time-orientation) is a member of $\obj(\frakM)$. Then the
injection map $\iota_{M,O}: (O,\gb_O) \to (M,\gb)$, i.e.\ the
identical map restricted to $O$, is an element in
$\Hom_{\frakM}((O,\gb_O),(M,\gb))$. With this notation, we obtain the
following assertion.
\begin{Pro} \label{localnet}
Let $\euA$ be a covariant functor with the properties stated in Def.\
\ref{CovFunctor}, 
and define a map $\calK(M,\gb) \owns O \mapsto \calA(O) \subset
\euA(M,\gb)$ by setting
$$ \calA(O) := \alpha_{M,O}(\euA(O,\gb_O))\,,$$
having abbreviated $\alpha_{M,O} \equiv \alpha_{\iota_{M,O}}$.
Then the following statements hold:
\begin{itemize}
\item[(a)] The map fulfills isotony, i.e.
 $$O_1 \subset O_2 \Rightarrow
\calA(O_1) \subset \calA(O_2) \ \mbox{for all}\  O_1,O_2 \in
\calK(M,\gb)\ .$$
\item[(b)] 
If there exists a group $G$ of isometric diffeomorphisms
$\kappa: M \to M$ (so that $\kappa_*\gb = \gb$) preserving orientation
and time-orientation, then there is a
representation $G \owns \kappa \mapsto \tilde{\alpha}_{\kappa}$ of $G$
by $C^*$-algebra automorphisms
$\tilde{\alpha}_{\kappa} : \calA \to \calA$ (where $\calA$ denotes the
minimal $C^*$-algebra generated by $\{\calA(O): O \in \calK(M,\gb)\}$)
such that
\begin{equation}
\label{cov1}
 \tilde{\alpha}_{\kappa}(\calA(O)) = \calA(\kappa(O))\,, \quad O \in
\calK(M,\gb)\,.
\end{equation}
\item[(c)] If the theory given by $\euA$ is additionally causal, then
  it holds that 
$$ [\calA(O_1),\calA(O_2)] = \{0\} $$
for all $O_1,O_2 \in \calK(M,\gb)$ with $O_1$ causally separated from
$O_2$.
\item[(d)] Suppose that the theory $\euA$ fulfills the time-slice
  axiom, and 
let $\Sigma$ be a Cauchy-surface in $(M,\gb)$ and let $S
  \subset \Sigma$ be open and connected. Then for each $O \in
  \calK(M,\gb)$ with $O \supset S$ it holds that
 $$ \calA(O) \supset \calA(S^{\perp}{}^{\perp}) $$
where $S^{\perp}{}^\perp$ is the double causal complement of $S$, and
$\calA(S^{\perp}{}^{\perp})$ is defined as the smallest $C^*$-algebra
formed by all $\calA(O_1)$, $O_1 \subset S^{\perp}{}^{\perp}$, $O_1
\in \calK(M,\gb)$.
\end{itemize}
\end{Pro} 
{\it Proof. }
$(a)$. The proof of this statement is based on the covariance properties of the
functor $\euA$. To demonstrate that isotony holds, let $O_1$ and $O_2$
be in $\calK(M,\gb)$ with $O_1 \subset O_2$. We denote by
$\iota_{2,1}: (O_1,\gb_{O_1}) \to (O_2,\gb_{O_2})$ the canonical
embedding obtained by restricting the identity map on $O_2$ to $O_1$,
hence $\iota_{2,1} \in
\Hom_{\frakM}((O_1,\gb_{O_1}),(O_2,\gb_{O_2}))$. With the notation
$\alpha_{\iota_{M,O_1}} \equiv \alpha_{M,1}$, etc., covariance of the
functor $\euA$ implies $\alpha_{M,1} = \alpha_{M,2} \circ
\alpha_{2,1}$ and therefore,
\begin{eqnarray*}
 & & \calA(O_1) = \alpha_{M,1}(\euA(O_1,\gb_{O_1})) =
 \alpha_{M,2}(\alpha_{2,1}(\euA(O_1),\gb_{O_1}))\\
& & \quad \quad \quad \quad \quad \quad \quad \subset
\alpha_{M,2}(\euA(O_2,\gb_{O_2})) = \calA(O_2)
\end{eqnarray*}
since $\alpha_{2,1}(\euA(O_1,\gb_{O_1})) \subset \euA(O_2,\gb_{O_2})$
by the very properties of the functor $\euA$.
\par
$(b)$. To prove the second part of the statement, let $\kappa: (M,\gb) \to
(M,\gb)$ be a diffeomorphism preserving the metric as well as
time-orientation and orientation. The functor assigns to it an
automorphism $\alpha_{\kappa}: \euA(M,\gb) \to \euA(M,\gb)$. Denoting
by $\tilde{\kappa}$ the map $O \to \kappa(O)$, $x \mapsto \kappa(x)$,
there is an associated morphism $\alpha_{\tilde{\kappa}}: \euA(O,\gb_O)
\to \euA(\kappa(O),\gb_{\kappa(O)})$. Hence we obtain the following
sequence of equations:
\begin{eqnarray*}
 \alpha_{\kappa}(\calA(O)) &=& \alpha_{\kappa} \circ
 \alpha_{M,O}(\euA(O,\gb_O)) = \alpha_{\kappa \circ
   \iota_{M,O}}(\euA(O,\gb_O))\\
& = & \alpha_{\iota_{M,\kappa(O)} \circ \tilde{\kappa}}(\euA(O,\gb_O)) 
 = \alpha_{M,\kappa(O)}\circ \alpha_{\tilde{\kappa}}(\euA(O,\gb_O))\\
& = & \alpha_{M,\kappa(O)}(\euA(\kappa(O),\gb_{\kappa(O)})) =
\calA(\kappa(O))\,.
\end{eqnarray*}
Since $\calA \subset \euA(M,\gb)$, it follows that defining
$\tilde{\alpha}_{\kappa}$ as the restriction of $\alpha_{\kappa}$ to
$\calA$ yields an automorphism with the required properties.
The group representation property is simply a consequence of the covariance
properties of the functor yielding $\alpha_{\kappa_1} \circ
\alpha_{\kappa_2} = \alpha_{\kappa_1 \circ \kappa_2}$ for any pair of
members $\kappa_1,\kappa_2 \in G$ together with \eqref{cov1} which
allows us to conclude that $\tilde{\alpha}_{\kappa_1} \circ
\tilde{\alpha}_{\kappa_2} = \tilde{\alpha}_{\kappa_1 \circ \kappa_2}$.
\par
$(c)$. If $O_1$ and $O_2$ are causally separated members in
$\calK(M,\gb)$, then one can find a Cauchy-surface $\Sigma$ in
$(M,\gb)$ and a pair of disjoint subsets $s_1$ and $S_2$ of $\Sigma$,
both of which are connected and relatively compact, so that $O_j
\subset S_j^{\perp}{}^\perp$, 
$j= 1,2$. Now $S_j^{\perp}{}^{\perp}$ are causally separated members
of $\calK(M,\gb)$, and  
equipped with the appropriate
restrictions of $\gb$ as metrics, they are globally hyperbolic spacetimes
in their own right, and naturally embedded into $(M,\gb)$. According
to the causally assumption on $\euA$, it holds that
$\calA(S_j^{\perp}{}^\perp)=
\alpha_{M,S_j^{\perp}{}^{\perp}}(\euA(S_j^{\perp}{}^{\perp}),\gb_{S_j^{\perp}
{}^\perp})$ 
are pairwise commuting subalgebras of $\euA(M,\gb)$, and due to
isotony, $\calA(O_j) \subset \calA(S_j^{\perp}{}^\perp)$, so that
$[\calA(O_1),\calA(O_2)] = \{0\}$.
\par
$(d)$. Consider $S^{\perp}{}^\perp$, equipped with the appropriate
restriction of $\gb$, as a globally hyperbolic spacetime in its own
right. Then $S$ is a Cauchy-surface for that spacetime, and $O\cap
S^{\perp}{}^\perp$ is an open neighbourhood of the Cauchy-surface
$S$. Hence there is an open neighbourhood $N$ of $S$
cointained in $O \cap S^{\perp}{}^\perp$ so that $N$, endowed with the
restricted metric, is again a globally hyperbolic spacetime. By the
time-slice axiom, it follows that
$\alpha_{S^{\perp}{}^\perp,N}(\euA(N)) = \euA(S^{\perp}{}^\perp)$,
where we have suppressed the metrics to ease notation. According to
the functorial properties of $\euA$ it follows that
$$ \calA(O) \supset \calA(N) = \calA(S^{\perp}{}^\perp)\,. $$
This completes the proof. ${}$ \hfill $\Box$ 
\\[6pt]
Thus, one can clearly see that, in the light of Prop.\ \ref{localnet}, the
Haag-Kastler framework is a special consequence of our functorial approach.

\subsection{Quantum Fields as Natural Transformations}
${}$\\[6pt]
We have just seen how a quantum field {\it theory} is defined in terms
of a covariant functor. 
There,  
an algebra is mapped via the endomorphism $\alpha_\psi$  into another 
algebra, but a priori there are no distiguished elements of the
algebras which are mapped onto each other by that transformation. It is
however useful to look  for such elements and it is actually what
motivated the whole approach. Indeed, we
shall look at the possibility to define locally covariant {\it fields}, and
their importance rests on the possibility to construct fields which,
in the light of our new principle of locality, depend only 
locally on the geometry.  
In a pair of interesting recent papers, Hollands and Wald \cite{HW,HW2} 
use this definition to
construct Wick-polynomials and time-ordered products of free scalar
fields as ``local and covariant fields,'' hence,
as objects  depending only locally on the metric.
In Sec.\ \ref{Sec5} we shall present  our own (but related) derivation
of Wick polynomials of a free scalar field, by solving a problem of
cohomological nature with a covariance constraint.

In a certain sense our definition gives rise to a locally 
covariant generalization of the
G{\aa}rding-Wightman approach to fields as operator-valued 
distributions. We here
do not insist at the beginning on having operators in a Hilbert space 
but, more abstractly, we consider them as distributions 
taking values in a topological *-algebra.  

The simplest definition may be given as follows: Consider a family 
$\Phi\equiv\{\Phi_{(M,\gb)}\}$,
indexed by all spacetimes $(M,\gb) \in \obj(\frakM)$, 
of quantum fields defined as 
``generalized algebra-valued distributions''.
That means, there is a family $\{\mathcal{A}(M,\gb)\}$ of topological
*-algebras  indexed by all spacetimes in $\obj(\frakM)$, 
and for each spacetime $(M,\gb)$,
$\Phi_{(M,\gb)} : C_0^{\infty}(M) \to \mathcal{A}(M,\gb)$ is a
continuous map (not necessarily linear, this is why we refer to it as
a ``generalized'' distribution).
 Consider in addition
any morphism $\psi\in\Hom_{\frakM}((M_1,\gb_1),(M_2,\gb_2))$. Then we
demand that there exists a continuous   
endomorphism $\alpha_\psi:\calA(M_1,\gb_1)\to\calA(M_2,\gb_2)$ 
so that, 
\begin{equation*}
\alpha_\psi(\Phi_{(M_1,\gb_1)}(f))=\Phi_{(M_2,\gb_2)}(\psi_*(f))
\end{equation*}
where $f\in C_0^{\infty}(M_1)$ is any test function 
and $\psi_*(f)=f\circ\psi^{-1}$ as before.
The family $\{\Phi_{(M,\gb)}\}$ with these covariance 
conditions is called a {\it locally
covariant} quantum field. This simple description has a beautiful
functorial translation, as we shall next outline.

We consider again the category $\frakM$, and introduce the category
$\frakTa$ consisting of topological *-algebras (with unit elements) 
as objects, and of
continuous *-endomorphisms as morphisms (i.e., $\alpha \in
\Hom_{\frakTa}(\mathcal{A}_1,\mathcal{A}_2)$ is a morphism of
$\frakTa$ if $\alpha: \mathcal{A}_1 \to \mathcal{A}_2$ is a
continuous, unit-preserving, injective *-morphism).
 In addition, we consider another category
$\fraktest$ which is the category containing as objects all 
possible test-function 
spaces over $\frakM$, that is, the objects consist of all spaces
$C_0^{\infty}(M)$ of smooth, compactly supported test-functions on $M$,
for $(M,\gb)$ ranging over the objects of $\frakM$, and   
the morphisms are all possible push-forwards $\psi_*$ of 
isometric embeddings $\psi: (M_1,\gb_1)\to (M_2,\gb_2)$. 
The action of any push-forward $\psi_*$ on an element of a 
test-function space has been defined above, and it clearly satisfies 
the requirements for morphisms between test-function spaces.

Now let a locally covariant
quantum field {\it theory} $\euA$  be defined as a functor
 in the same manner as in Def.\ \ref{CovFunctor}, but with the
 category $\frakTa$ in place of the category $\frakA$, and again
 following the convention to denote $\euA(\psi)$ by $\alpha_{\psi}$ whenever 
$\psi$ is any morphism in $\frakM$. Moreover, let $\euD$ be the covariant
functor between $\frakM$ and $\fraktest$  assigning to
each $(M,\gb) \in \obj(\frakM)$ the test-function space $\euD(M,\gb) =
C_0^{\infty}(M)$, and to each morphism $\psi$ of $\frakM$ its
push-forward: $\euD(\psi) = \psi_*$. 
We regard the categories $\fraktest$ and $\frakTa$
as subcategories of the category of all topological spaces
$\fraktopvec$, and hence we are led to adopt the following
\begin{Def} \label{nattrans} ${}$\\
 A {\bf locally covariant quantum field} $\Phi$ is a
natural transformation between the functors $\euD$ 
and $\euA$, i.e.\ for any object
$(M,\gb)$ in $\frakM$ there exists a morphism 
$\Phi_{(M,\gb)}:\euD(M,\gb)\to
\euA(M,\gb)$ in $\fraktopvec$  such that  for  
each given morphism\\ $\psi \in \Hom_{\frakM}((M_1,\gb_1),(M_2,\gb_2))$ 
the following diagram
\begin{equation*}
\begin{CD}
\euD(M_1,\gb_1) @>\Phi_{(M_1,\gb_1)}>> \euA(M_1,\gb_1)\\
@V{\psi_*}VV     @VV{\alpha_{\psi}}V\\
\euD(M_2,\gb_2)@>>\Phi_{(M_2,\gb_2)}> \euA(M_2,\gb_2)
\end{CD}
\end{equation*}
commutes. 
\end{Def}

The commutativity of the diagram means, explicitly, that 
\begin{equation*}
\alpha_\psi\circ\Phi_{(M_1,\gb_1)} = \Phi_{(M_2,\gb_2)}\circ \psi_*
\end{equation*}
i.e., the requirement of covariance for fields. 
\\[6pt]
\noindent{\it Remarks.} 
(A) This definition may of course be extended; instead of the
test-function spaces $C_0^\infty(M)$ one may take smooth compactly
supported sections of vector bundles, and endomorphisms of such more
general test-sections spaces which are suitable pull-backs of
vector-bundle endomorphisms. Also, one might include conditions on the
wave-front set of the field-operators.
\\[6pt]
(B) The notion of causality  may also be introduced
in the obvious manner: One calls a locally covariant quantum field {\it
  causal} if for all $f,h  \in \euD(M,\gb)$ it holds that
$\Phi_{(M,\gb)}(f)$ and $\Phi_{(M,\gb)}(h)$ commute. 
\\[6pt]
(C) One reason for allowing non-linear fields in the definitions of
quantum fields as natural transformations is that it can be applied to
more general objects. One would be the definition of a locally
covariant $S$-matrix, patterned after the definition of a the
``local'' $S$-amtrix of Epstein and Glaser, see e.g.\ \cite{BF}. At
the perturbative level (in the sense of formal power series) this
amounts to showing that time-ordered products may be defined in such a
way that they become locally covariant fields. Indeed, in a recent
paper \cite{HW2}, Hollands and Wald successfully proved the existence
of such locally covariant, time-ordered fields. At the
non-perturbative level, it might be possible that the constraint of
local covariance together with a dynamical generator property (in the
spirit of Sec.\ 4) allows it to fix the phase of the $S$-matrix. We
hope to return elsewhere to this issue. 
\subsection{Free Scalar Klein-Gordon Field as a Natural
  Transformation} ${}$ \\[6pt]
The present subsection serves the purpose of sketching two simple
examples for locally covariant quantum fields. The first example is
based on the Borchers-Uhlmann algebra which can be associated with
each manifold $M$. It assigns to each differentiable manifold $M$ a
topological *-algebra $\mathfrak{B}(M)$ that is constructed as follows:
Elements in $\mathfrak{B}(M)$ are sequences $(f_n)$ $(n \in
\mathbb{N}_0)$ where $f_0 \in \mathbb{C}$ and $f_n \in C_0^\infty(M^n)$ for $n
> 0$. Addition and scalar multiplication are defined as usual for
sequences with values in vector spaces, and the product $(f_n)(h_n)$ in
$\mathfrak{B}(M)$ is defined as the sequence $(j_n)$ where
 $$ j_n(x_1,\ldots,x_n) = \sum_{i+j =
   n}f_i(x_1,\ldots,x_i)h_j(x_{i+1},\ldots,x_n)\,, \quad
 (x_1,\ldots,x_n) \in M^n\,.$$
The *-operation is defined via $(f_n)^* = (\overset{=}{f}{}_n)$ where
$\overset{=}{f}{}_n(x_1,\ldots,x_n) = \overline{f_n(x_n,\ldots,x_1)}$,
the latter overlining meaning complex conjugation. The unit element is
given by ${\bf 1} = (1,0,0,\ldots)$. The algebra can be
equipped with a fairly natural locally convex topology with respect to
which it is complete. See \cite{Bor}, \cite{Uhl} (and also \cite{FH},
\cite{SV1} in the context of curved spacetime manifolds)
 for further discussion of the Borchers-Uhlmann algebra.

Given an endomorphism $\psi \in
\Hom_{\frakM}((M_1,\gb_1),(M_2,\gb_2))$, one can lift it to an
algebraic endomorphism $\alpha_{\psi}: \mathfrak{B}(M_1) \to
\mathfrak{B}(M_2)$ by setting
$$ \alpha_{\psi}((f_n)) = (\psi_*^{(n)}f_n) $$
where $\psi^{(n)}_*$ denotes the $n$-fold push-forward, given by
$(\psi_*^{(n)}f_n)(y_1,\ldots,y_n) =$ \\
$f_n(\psi^{-1}(y_1),\ldots,\psi^{-1}(y_n))$. We thus obtain a
covariant functor $\euA$ between $\frakM$ and $\frakTa$ by setting
$\euA(M,\gb)= \mathfrak{B}(M)$ and $\euA(\psi) = \alpha_{\psi}$ as
just defined.
A locally covariant quantum field $\Phi$ in the sense of Def.\
\ref{nattrans} may then be obtained by defining for $(M,\gb) \in
\obj(\frakM)$ and $f \in \euD(M,\gb) = C_0^{\infty}(M)$,
$$ \Phi_{(M,\gb)}(f) = (f_n) $$
where $(f_n) \in \euA(M,\gb) = \mathfrak{B}(M)$ is the sequence with
$f_1 = f$ and $f_n = 0$ for all $n \ne 1$. It is straightforward to
check that this indeed satisfies all conditions for a natural
transformation with respect to the functors $\euD$ and $\euA$.
\\[6pt]
The Borchers-Uhlmann algebra, however, carries no dynamical
information, which would have to be incorporated by passing to
representations, or factorizing by ideals. In this spirit, we
introduce as our second example the Klein-Gordon field as a locally
covariant field. For $(M,\gb) \in \obj(\frakM)$, let $J(M,\gb)$ be the
(closed) two-sided ideal in $\mathfrak{B}(M)$ that is generated by all
the terms
$$ (f_n)(h_n) - (h_n)(f_n) - \sigma(Ef,Eh){\bf 1} $$
where the $(f_n)$ and $(h_n)$ in $\mathfrak{B}(M)$ are such that $f_1
= f$, $h_1 = h$, and all other entries in the sequences vanish; $E =
E_{(M,\gb)}$ and $\sigma = \sigma_{(M,\gb)}$ are the propagator and
symplectic form corresponding to the Klein-Gordon equation 
\begin{equation} \label{KG22}
 (\nabla^{\mu}\nabla_{\mu} + \xi R + m^2)\varphi = 0
\end{equation}
on $(M,\gb)$ introduced in Subsection \ref{kgfe}. (Again it is assumed
that the constants $\xi$ and $m$ are the same for all $(M,\gb)$).

Then we introduce a new functor $\euA$ between $\frakM$ and $\frakTa$,
as follows: We define $\euA(M,\gb) = \mathfrak{B}(M)/J(M,\gb)$ and,
denoting by $[\,.\,]: \mathfrak{B}(M) \to \mathfrak{B}(M)/J(M,\gb)$
the quotient map, we set for $\psi \in
\Hom_{\frakM}((M_1,\gb_1),(M_2,\gb_2))$,
$$ \euA(\psi)([(f_n)]) \equiv \alpha_{\psi}([(f_n)]) =
[(\psi_*^{(n)}f_n)]$$
where $\psi^{(n)}_*$ is the $n$-fold push-forward of $\psi$ defined
above. The required properties of this definition of $\alpha_{\psi}$
to map $J(M_1,\gb_1)$ into $J(M_2,\gb_2)$, and $\alpha_{\psi \circ
  \psi'} = \alpha_{\psi} \circ \alpha_{\psi'}$, can be obtained by an
argument similar to that in Subsection \ref{kgfe} showing that the
$\alpha_{\psi}$ defined there have the desired covariance properties.

With respect to this new functor $\euA$, we may now define the
generally covariant Klein-Gordon field $\Phi$ as a natural
transformation according to Def.\ \ref{nattrans} through setting for
$(M,\gb) \in \obj(M,\gb)$ and $f \in \euD(M,\gb) = C_0^{\infty}(M)$,
$$ \Phi_{(M,\gb)}(f) = [(f_n)]$$
where, as above, $(f_n)$ is the element in $\mathfrak{B}(M)$ with $f_1
= f$ and $f_n = 0$ for all $n \ne 1$. Again, the properties of a
natural transformation are easily checked for this definition.

Moreover, locally covariant quantum fields $\Phi$ modelling the
Klein-Gordon field \eqref{KG22} may be obtained from the functor
$\euA$ of Subsection \ref{kgfe} describing the locally covariant
quantum field theory of the Klein-Gordon field at $C^*$-algebraic
level. We give only a rough sketch of the idea. Let $\euA$ be the
functor associated with the Klein-Gordon field in Subsection
\ref{kgfe}. Let $(M,\gb) \in \obj(\frakM)$, and let $\pi$ be a
Hilbert-space representation of the $C^*$-algebra $\euA(M,\gb)$ on a
representation Hilbert-space $\mathcal{H}$. We assume that 
there exists a dense subspace $\mathcal{V}$ of $\mathcal{H}$ so that,
for each $f \in C_0^{\infty}(M,\R)$, the field operator
$$ \Phi_{(M,\gb)}(f) = \left. \frac{d}{ds}\right|_{s=0} \pi(W(sEf))$$
exists as an (essentially) 
self-adjoint operator on $\mathcal{V}$, where $E$ denotes
the propagator and $W(\,.\,)$ the Weyl-algebra generators associated
with the Klein-Gordon field on $(M,\gb)$. (The field operators can be
extended to all complex-valued testfunctions by requiring complex
linearity.) The notation used here
already suggests how one may go about in order to try to obtain a
locally covariant quantum field in this way. Supposing a quantum field
$\Phi_{(M,\gb)}$ can be defined in this manner for all $(M,\gb) \in
\obj(\frakM)$ (from representations $\pi$ for each spacetime), and
that, for each $\psi \in \Hom_{\frakM}((M,\gb),(M',\gb'))$, the assignment 
$\tilde{\alpha}_{\psi}(\Phi_{(M,\gb)}(f)) = \Phi_{(M',\gb')}(\psi_*f)$
extends to a *-algebraic endomorphism $\tilde{\alpha}_{\psi}:
\tilde{\euA}(M,\gb) \to \tilde{\euA}(M',\gb')$, where $\tilde{\euA}(M,\gb)$
denotes the *-algebra formed by all the $\Phi_{(M,\gb)}(f)$, $f \in
C_0^\infty(M)$, one obtains in this way a locally covariant quantum
field $\Phi$ as a natural transformation.

In a similar spirit, Hollands and Wald have constructed Wick-ordered
and time-ordered products of the free scalar Klein-Gordon field, starting
from quasifree Hadamard representations, such that these product-fields are
locally covariant fields and are natural transformations in the sense
of Def.\ \ref{nattrans} \cite{HW,HW2}. We refer to these references
for further discussion, and also to Sec.\ \ref{Sec5}.
\section{States, Representations, and the Principle of Local
  Definiteness}
\subsection{Functorial Description of a State Space} \label{FDSS}
${}$\\[6pt]
The description of a physical
system in terms of operator algebras requires also the concept of
states so that expectation values of observables can be
calculated. First, suppose that one is given  a
$C^*$-algebra $\calA$ with unit element ${\bf 1}$
modelling the algebra of observables of some physical system. 
A state is a linear functional $\omega : \calA \to \mathbb{C}$
having the property of being positive,
i.e.\ $\omega(A^*A) \ge 0$ $\forall A \in \calA$, and normalized,
i.e.\ $\omega({\bf 1}) = 1$.
Thus, given any hermitean element $A \in \calA$, the number
$\omega(A)$ is interpreted as an expectation value of the observable
$A$ in the state $\omega$.

 There is an intimate relation between states on $\calA$ and Hilbert-space
representations of $\calA$. If $\pi$ is a linear $*$-representation of
$\calA$ by bounded linear operators on some Hilbert-space $\calH$, then each
positive density matrix $\rho$ with unit trace on $\calH$ induces a
state $\omega(A) = {\rm tr}(\rho\cdot \pi(A))$, $a \in \calA$, on
$\calA$. There is also a converse of that: For each state $\omega$ on
$\calA$ there exists a triple
$(\calH_{\omega},\pi_{\omega},\Omega_{\omega})$, consisting of a
Hilbert-space $\calH_{\omega}$, a linear $*$-representation
$\pi_{\omega}$ of $\calA$ by bounded linear operators on
$\calH_{\omega}$, and a unit vector $\Omega_{\omega} \in
\calH_{\omega}$ such that $\omega(A) = \langle
\Omega_{\omega},\pi_{\omega}(A)\Omega_{\omega}\rangle$ for all $A \in
\calA$. This triple is called the GNS-representation of $\omega$
(after Gelfand, Naimark and Segal); for its construction, see e.g.\
\cite{BR1}.

Now suppose that our set of observables arises in terms of a functor
$\euA$ describing a locally covariant
quantum field theory. The question arises what the concept of a state
might be in this case. The first, quite natural idea is to say that a
state is a family $\{\omega_{(M,\gb)}: (M,\gb) \in \obj(\frakM) \}$
indexed by the members in the object-class $\frakM$ where each
$\omega_{(M,\gb)}$ is a state on the $C^*$-algebra
$\euA(M,\gb)$. Usually, however, one is interested in states with
particular properties, e.g., one would like to consider states
$\omega_{(M,\gb)}$ fulfilling an appropriate variant of the
``microlocal spectrum condition'' \cite{BFK} which can be seen as a
replacement for the relativistic spectrum condition for quantum field
theories on curved spacetime and which, for free fields, is equivalent
to the Hadamard condition (cf.\ Sec.\ \ref{kgfe}, and \cite{Rad1,SV2}). One
might wonder if, above that, there are families of states
$\{\omega_{(M,\gb)}: (M,\gb) \in \obj(\frakM)\}$ that are
distinguished by a property which in our framework would correspond to
``local diffeomorphism invariance'', namely,
$$ \omega_{(M',\gb')} \circ \alpha_{\psi} = \omega_{(M,\gb)}\quad {\rm
  on}\ \ \euA(M,g)$$
for all $\psi \in \Hom_{\frakM}((M,\gb),(M',\gb'))$. However, it has been shown
in \cite{HW} that this invariance property cannot be realized for
states of the free scalar field fulfilling the microlocal spectrum
condition. Let us briefly sketch an argument showing that the above
property will, in general, not be physically realistic.
 Let us consider two spacetimes $(M_1,\gb_1)$
and $(M_2,\gb_2)$, and assume that $(M_1,\gb_1)$ is just
Minkowski-spacetime. Moreover, it will be assumed that $(M_2,\gb_2)$
consists of three regions which are themselves globally hyperbolic
sub-spacetimes of $(M_2,\gb_2)$: An ``intermediate'' region $L_2$
lying to the future of a region $N_2^-$ and to the past of a region
$N_2^+$. All these regions are assumed to contain Cauchy-surfaces, and
it is also assumed that the regions $N_2^{\pm}$ are isometrically
diffeomorphic to globally hyperbolic subregions $N_1^{\pm}$ of
Minkowski spacetime $(M_1,\gb_1)$ which likewise contain
Cauchy-surfaces. By $\iota^{\pm}:N_1^{\pm} \to N_2^{\pm}$ we denote the
corresponding isometric diffeomorphisms. We may, for the sake of
concreteness, consider a free scalar field (cf.\ next section), and
define the state $\omega_1$ on $\euA(M_1,\gb_1)$ to be its vacuum
state (which fulfillys the microlocal spectrum condition). Then the 
state $\omega_2^- = \omega_1 \circ \alpha_{\iota^-}^{-1}$ induces a
state on $\euA(N_2^-,\gb_{2,N_2^-})$ and thereby, since the free field
obeys the time-slice axiom, it induces a state $\omega_2$ on
$\euA(M_2,\gb_2)$ (which again fulfills the microlocal spectrum condition). 
Now the state $\omega_2$ restricts to a state
$\omega_2^+$ on $\euA(N_2^+,\gb_{2,N^+_2})$. However, if there is
non-trivial curvature in the intermediate region $L_2$, then the state
$\omega_2$, which was a vacuum state on the ``initial'' region
$N_2^-$, will no longer be a vacuum state on the ``final'' region
$N_2^+$ \cite{Wald.scat}.
 The regions $N_2^-$ and $N^+_2$ possess isometric subregions;
it is no loss of generality to suppose that there is an isometric
diffeomorphism $\psi: N_2^- \to N_2^+$. Then invariance in the above
sense  of the family
of states $\omega_1,\omega_2,\omega_2^{\pm}$  demands that 
$$ \omega_2^+ \circ \alpha_{\psi} = \omega_2^-\,,$$
but this is not the case ($\omega_2^-$ is (the restriction of) a vacuum
state, $\omega_2^+$ is (the restriction of) a non-vacuum state.) 
The counterexample is based on a form of ``relative
Cauchy-evolution'', which is worth being studied in greater
generality, and this will be the topic of Section \ref{Dynamics}.

In view of this negative result one finds oneself confronted with the
question if there is a more general concept of ``invariance'' that can
be attributed to families of states
$\{\omega_{(M,\gb)}:(M,\gb) \in \obj(\frakM)\}$ for a locally
covariant quantum field theory given by a
functor $\euA$. We will argue that there is a positive answer to that
question: The local folia determined by states satisfying the
microlocal spectrum condition are good candidates  for minimal classes
of states which are locally diffeomorphism covariant. To explain this,
let us fix some concepts.
\\[6pt]
{\bf Folium of a representation.} 
 Let $\mathcal{A}$ be a $C^*$-algebra and $\pi: \mathcal{A} \to
 B(\calH)$ a $*$-representation of $\mathcal{A}$ by bounded linear
 operators on a Hilbert space $\calH$. The {\it folium} of $\pi$,
 denoted by $\boldsymbol{F}(\pi)$, is the set of all states $\omega'$
 on $\mathcal{A}$ which can be written as 
$$ \omega'(A) = {\rm tr}(\rho \cdot \pi(A))\,, \quad A \in
\euA(M,\gb)\ .$$
 In other words, the folium of a representation consists of all
 density matrix states in that representation.
\\[6pt]
{\bf Local quasi-equivalence and local normality.}
Let $\euA$ be a locally covariant quantum
field theory and let, for $(M,\gb)$ fixed, $\omega$ and $\tilde{\omega}$ 
be two states on $\euA$. We will say that these
states (or their GNS-representations, denoted by $\pi$ and
$\tilde{\pi}$, respectively) are {\it locally
  quasi-equivalent} if for all
$O \in \calK(M,\gb)$ the relation
\begin{equation} \label{F1}
 \boldsymbol{F}(\pi \circ \alpha_{M,O}) =
\boldsymbol{F}(\tilde{\pi} \circ \alpha_{M,O})
\end{equation}
is valid, where $\alpha_{M,O} = \alpha_{\iota_{M,O}}$ and
$\iota_{M,O}: (O,\gb_O) \to (M,\gb)$ is the natural embedding (cf.\
Prop.\ \ref{localnet}).

Moreover, we say that $\omega$ is {\it locally normal} to $\tilde{\omega}$
(or to the corresponding GNS-representation $\tilde{\pi}$) if 
\begin{equation}
\label{F2}
 \omega \circ \alpha_{M,O} \in \boldsymbol{F}(\tilde{\pi} \circ
 \alpha_{M,O})
\end{equation}
holds for all $O \in \calK(M,\gb)$.
\\[6pt]
{\bf Intermediate factoriality.} 
Let $\omega$ be a state on $\euA(M,\gb)$, then we define for each $O
\in \calK(M,\gb)$ the von Neumann algebra
$\mathcal{M}_{\omega}(O) = \pi_{\omega}(\alpha_{M,O}(\euA(M,\gb)))''$,
the local von Neumann algebra of the region $O$ with respect to the
state $\omega$. We say that the state $\omega$ fulfills the condition
of {\it intermediate factoriality} if for each $O \in \calK(M,\gb)$
there exist $O_1 \in \calK(M,\gb)$ and a factorial von Neumann algebra
$\mathcal{N}$ acting on the GNS-Hilbert-space $\mathcal{H}_{\omega}$
of $\omega$ so that 
$$ \mathcal{M}_{\omega}(O) \subset \mathcal{N} \subset
\mathcal{M}_{\omega}(O_1)\,. $$
(We recall that a factorial von Neumann algebra $\mathcal{N}$ is a
von Neumann algebra so that $\mathcal{N} \cap \mathcal{N}'$ contains
only multiples of the unit operator.)
\\[6pt]
It is known that quasifree states of the free scalar field on globally
hyperbolic spacetimes which fulfill the microlocal spectrum condition
have the property to be locally quasi-equivalent (cf.\ Subsec.\
\ref{3.2}). Thus, local quasi-equivalence may be expected for states
satisfying the microlocal spectrum condition.
More generally, local normality can be interpreted as ruling out the
possibility of local superselection
rules. Also intermediate factoriality is known to hold for states of
the free scalar field fulfilling the microlocal spectrum condition on
globally hyperbolic spacetimes (cf.\ again Sec.\ 3). The condition of
intermediate factoriality serves the purpose of eliminating the possible
difference between the folium of a representation and the folium of
any of its (non-trivial) subrepresentations (see Appendix {\bf b)}). It
can also be motivated as the consequence of a stricter formulation,
known a ``split property'', which is expected to hold for all (also
interacting) physically relevant quantum field theories on general
grounds (cf.\ \cite{Sum.Rev,Haag,BuWich}) and is in fact known to hold
for states of the free field 
fulfilling the microlocal spectrum condition in flat and curved
spacetimes \cite{Bu.Split,Ver3}, and for interacting theories in low
dimensions \cite{Sum.Split}.
 We also note that the property
of a state to fulfill the microlocal spectrum condition is a locally
covariant  property (owing to the covariant behaviour of
wavefront sets of distributions under diffeomorphisms \cite{Horm})
 and thus, for a
locally covariant quantum field theory it is natural to
assume that, if $\omega_{(M',\gb')}$ fulfills (any suitable variant of)
the microlocal spectrum condition, then so does
$\omega_{(M',\gb')}\circ \alpha_{\psi}$ for any $\psi \in
\Hom_{\frakM}((M,\gb),(M',\gb'))$. In the case where
 also the folia of states (i.e.,
the folia of their GNS-representations)
satisfying the microlocal spectrum condition coincide locally, one thus
obtains the invariance of local folia under local diffeomorphisms for
families of states satisfying the microlocal spectrum condition, more
precisely, at the level of the GNS-representations of
$\omega_{(M,\gb)}$ and $\omega_{(M',\gb')}$,
$$
\boldsymbol{F}(\pi_{(M',\gb')}\circ \alpha_{\psi} \circ
\alpha_{M,O}) = \boldsymbol{F}(\pi_{(M,\gb)} \circ \alpha_{M,O})
$$
holds for all $\psi \in \Hom_{\frakM}((M,\gb),(M',\gb'))$ and all $O \in
\calK(M,\gb)$. All these properties are known to hold for quasifree
states of the free scalar field fulfilling the microlocal spectrum
condition on global hyperbolic spacetimes, see Subsec.\ \ref{3.2} for
 discussion.

Thus one can see that local diffeomorphism invariance really occurs at
the level of local folia of states for $\euA$. In this light, it
appears natural to give a functorial description of the space of
states that takes this form of local diffeomorphism invariance into
account.
To this end, it seems convenient to first introduce a new category,
the category of the set of states.
\begin{description}
\item[$\frakS$] An object $\boldsymbol{S}\in \obj(\frakS)$ is a 
  set of states on a $C^*$-algebra $\mathcal{A}$.
 Morphisms between members $\boldsymbol{S}'$ and
  $\boldsymbol{S}$ of $\obj(\frakS)$ are positive maps $\gamma^* :
  \boldsymbol{S}' \to \boldsymbol{S}$. In the present work,
  $\gamma^*$ arises always as the dual map of a faithful $C^*$-algebraic
  endomorphism $\gamma: \mathcal{A} \to \mathcal{A}'$ via
$$
 \gamma^* \omega'(A) = \omega'(\gamma(A))\,, \quad \omega' \in
 \boldsymbol{S}',\  \ A \in \mathcal{A}\,.
$$
The category $\frakS$ is therefore ``dual'' to the category $\frakA$.
The composition rules for morphisms should thus be obvious.
\end{description}
Now we can define a state space for a locally covariant
quantum field theory in a functorial manner.
\begin{Def}${}$\\
Let $\euA$ be a locally covariant quantum
field theory.\\[4pt]
$(i)$\quad
A {\bf state space} for $\euA$ is a contravariant functor $\boldsymbol{S}$ 
between $\frakM$ and $\frakS$:
\begin{equation*}
\begin{CD}
(M,\gb) @>\psi>> (M',\gb')\\
@V{\boldsymbol{S}}VV     @VV{\boldsymbol{S}}V\\
\boldsymbol{S}(M,\gb)@<{\alpha_\psi^*}<< \boldsymbol{S}(M',\gb')
\end{CD}
\end{equation*}
where $\boldsymbol{S}(M,\gb)$ is a set of states on $\euA(M,\gb)$ and
$\alpha_{\psi}^*$ is the dual map of $\alpha_{\psi}$;
the contravariance property is
$$ \alpha^*_{\tilde{\psi} \circ \psi} = \alpha^*_{\psi} \circ
\alpha^*_{\tilde{\psi}} $$
together with the requirement that unit morphisms are mapped to unit morphisms.
\\[4pt]
$(ii)$ \quad 
We say that a state space $\boldsymbol{S}$ is {\bf locally quasi-equivalent} if
  eqn.\ \eqref{F1} holds for any pair of states $\omega,
  \tilde{\omega}  \in
  \boldsymbol{S}(M,\gb)$ (with GNS-representations $\pi,\tilde{\pi}$) 
whenever $(M,\gb)$ and $O \in \calK(M,\gb)$.
\\[4pt]
$(iii)$ \quad
A state space $\boldsymbol{S}$ is called {\bf locally normal} if
there exists a locally
quasi-equivalent state space $\tilde{\boldsymbol{S}}$ so that for each
$\omega \in \boldsymbol{S}(M,\gb)$ there is some $\tilde{\omega} \in
\tilde{\boldsymbol{S}}(M,\gb)$ (with GNS-representation $\tilde{\pi}$) 
so that \eqref{F2} holds for all $O \in
\calK(M,\gb)$.
\\[4pt] 
$(iv)$ \quad We say that a state space $\boldsymbol{S}$ 
is {\bf intermediate factorial} if each
state $\omega \in \boldsymbol{S}(M,\gb)$ fulfills the condition of
intermediate factoriality.  
\end{Def}
We list a few direct consequences of the previous definitions.
\begin{Thm} \label{ThmFolia}
${}$ \\
$(a)$ \quad Let $\boldsymbol{S}$ be a state space which is
intermediate factorial. Then
for all spacetimes $(M,\gb), (M',\gb') \in \obj(\frakM)$ and all pairs of
states $\omega \in \boldsymbol{S}(M,\gb)$, $\omega' \in
\boldsymbol{S}(M',\gb')$ with GNS-re\-pre\-sen\-tations
 $\pi$, $\pi'$ there holds
\begin{equation} \label{foltrans}
 \boldsymbol{F}(\pi' \circ \alpha_{\psi} \circ \alpha_{M,O}) =
 \boldsymbol{F}(\pi \circ \alpha_{M,O})\,, \quad O \in \calK(M,\gb)\,,
\end{equation}
if and only if the state space is locally quasi-equivalent.
\\[4pt]
$(b)$ \quad
If the state space $\boldsymbol{S}$ is locally normal, then there
exists a family of states $\{\omega_{(M,\gb)}:(M,\gb) \in
\obj(\frakM)\}$ on $\euA$ with the property that each $\omega \in
\boldsymbol{S}(M,\gb)$ is locally normal to $\omega_{(M,\gb)}$.
\\[4pt]
$(c)$ \quad If $\tilde{\boldsymbol{S}}$ is a locally quasi-equivalent
and intermediate factorial
state space, then one obtains a convex, locally normal state space
$\boldsymbol{S}$  by
defining $\boldsymbol{S}(M,\gb)$ as the set of all states which are
locally normal to any state on $\tilde{\boldsymbol{S}}(M,\gb)$. 
\end{Thm}
\noindent 
{\it Proof.}
In our proof, we will make use of the following statements:
\begin{itemize}
\item[$(\alpha)$] Let $\mathcal{A},\mathcal{B}$ and $\mathcal{C}$ be
  $C^*$-algebras with $C^*$-algebraic endomorphisms
$$ \mathcal{A} \overset{\beta}{\longrightarrow} \mathcal{B}
\overset{\gamma}{\longrightarrow} \mathcal{C}\,,$$
and let $\omega$ be a state on $\mathcal{C}$. Then there holds
$$ \boldsymbol{F}(\pi_{\omega} \circ \gamma \circ \beta) \supset
\boldsymbol{F}(\pi_{\omega \circ \gamma} \circ \beta) \supset
\boldsymbol{F}(\pi_{\omega \circ \gamma \circ \beta})\,,$$
where $\pi_{\nu}$ denotes the GNS-representation of the state $\nu$,
we will use this notation also below.
\item[$(\beta)$] Let $\mathcal{N}$ be a factorial von Neumann algebra on
  some Hilbert-space $\mathcal{H}$, and let $\mathcal{H}_{\mathcal N}$
  be some $\mathcal{N}$-invariant closed, non-zero subspace. Then for
  every density matrix $\rho = \sum_i \lambda_i |\phi_i\rangle \langle
  \phi_i|$, where the $\phi_i$ are unit vectors in $\mathcal{H}$,
  there exists a density matrix $\rho^{\mathcal N} = \sum_j \mu_j |
  \chi_j\rangle \langle \chi_j|$, where the $\chi_j$ are unit vectors
  in $\mathcal{H}_{\mathcal N}$, so that
\begin{equation} \label{trN}
 {\rm tr}(\rho \cdot N) = {\rm tr}(\rho^{\mathcal N} \cdot N)
\end{equation} 
holds for all $N \in \mathcal{N}$.
\end{itemize}
These statements will be proved in the appendix.
\\[4pt]
$(a)$. A first immediate observation is that
$\alpha^*_{\psi}\boldsymbol{S}(M',\gb') \subset \boldsymbol{S}(M,\gb)$
together with the condition of local quasi-equivalence imply 
\begin{equation} \label{foleq}
\boldsymbol{F}(\pi_{\omega' \circ \alpha_{\psi}} \circ \alpha_{M,O} )
= \boldsymbol{F}(\pi \circ \alpha_{M,O}) \,, \quad O \in
\calK(M,\gb)\,.
\end{equation}
Now fix $O \in \calK(M,\gb)$. According to the assumed condition of
intermediate factoriality, there are a region $O_1 \in
\calK(M,\gb)$ and a factorial von Neumann algebra $\mathcal{N}$ so
that
$$ \mathcal{M}_{\omega'}(\psi(O)) \subset \mathcal{N} \subset
\mathcal{M}_{\omega'}(\psi(O_1))\,.$$
Consequently, if we choose an arbitrary state $\omega_1 \in
\boldsymbol{F}(\pi_{\omega'} \circ \alpha_{\psi}\circ \alpha_{M,O})$,
then there exists, according to statement $(\beta)$ above, a density
matrix $\rho^{\mathcal N} = \sum_j \mu_j |\chi_j\rangle \langle
\chi_j|$ with $\chi_j \in \mathcal{H}_{\mathcal N} =
\overline{\mathcal{N}\Omega'}$ (where $\Omega'$ is the GNS-vector of
$\omega'$) with the property 
$$\omega_1(A) = {\rm tr}(\rho^{\mathcal N}\cdot \pi_{\omega'}\circ
\alpha_{\psi}\circ \alpha_{M,O}(A))\,, \quad A \in
\alpha_{M,O}(\euA(O,\gb_O))\,.    $$
Therefore, the state is in particular given by a density matrix
$\rho^{\mathcal{N}}$ in the GNS-representation of $\omega' \circ
\alpha_{M',\psi(O_1)}$, so that $\omega_1$ extends to a state    
$$\overline{\omega}_1 \in \boldsymbol{F}(\pi_{\omega' \circ
  \alpha_{M',\psi(O_1)}})\,.  $$
Owing to covariance, this in turn shows that
 $$\overline{\omega}_1
\in \boldsymbol{F}(\pi_{\omega' \circ \alpha_{\psi} \circ
\alpha_{M,O_1}})\,.$$ 
Restricting $\overline{\omega}_1$ again to $\omega_1
= \overline{\omega}_1 \circ \alpha_{M,O}$ on $\euA(O,\gb_O)$ yields
 $$ \omega_1 \in \boldsymbol{F}(\pi_{\omega' \circ \alpha_{\psi}} \circ
 \alpha_{M,O})\,.$$
In view of statement $(\alpha)$ above and because of \eqref{foleq}, we
have thus shown that \eqref{foltrans} holds for all $O \in
\calK(M,\gb)$ if $\boldsymbol{S}$ is locally quasi-equivalent. The
reverse implication, saying that \eqref{foltrans} implies that
$\boldsymbol{S}$ is locally quasi-equivalent, is evident.
\\[4pt]
$(b)$. One may choose an arbitrary family of states $\omega_{(M,\gb)}
\in \tilde{\boldsymbol{S}}(M,\gb)$; since each such choice of states is locally
quasi-equivalent to any other, by definition each state in
$\boldsymbol{S}(M,\gb)$ will be locally normal to $\omega_{(M,\gb)}$.
\\[4pt]
$(c)$. If $\boldsymbol{S}$ is a state space, then it is clearly locally
normal owing to the way it is defined. So it suffices to prove that
$\boldsymbol{S}$ is a state space, and convex.

To show that $\boldsymbol{S}$ is a state space, it is enough to
demontrate that
$$ \alpha^*_{\psi}(\boldsymbol{S}(M',\gb')) \subset
\boldsymbol{S}(M,\gb)\,, $$
since the contravariance property of the $\alpha^*_{\psi}$'s is
inherited from the covariance property of the $\alpha_{\psi}$'s.
Now if $\omega' \in \boldsymbol{S}(M',\gb')$, then this means that
$$ \omega'\circ \alpha_{M',O'} \in \boldsymbol{F}(\pi_{\hat{\omega}} \circ
\alpha_{M',O'})$$
holds for all $O' \in \calK(M',\gb')$, where $\hat{\omega}$ is some
element in $\tilde{\boldsymbol{S}}(M',\gb')$. Using covariance one
deduces from this relation
$$ (\alpha_{\psi}^*\omega') \circ \alpha_{M,O} = \omega' \circ
\alpha_{\psi} \circ \alpha_{M,O} \in
\boldsymbol{F}(\pi_{\hat{\omega}} \circ \alpha_{\psi} \circ
\alpha_{M,O})\,.$$
Then part $(a)$ of the proposition entails
$$ (\alpha_{\psi}^*\omega') \circ \alpha_{M,O} \in
\boldsymbol{F}(\pi_{\tilde{\omega}} \circ \alpha_{M,O})$$
for all $O \in \calK(M,\gb)$ with some $\tilde{\omega} \in
\tilde{\boldsymbol S}(M,\gb)$, showing that $\alpha^*_{\psi}\omega' \in
\boldsymbol{S}(M,\gb)$. 

Finally, we show that $\boldsymbol{S}$ is convex. Let
$\omega' = \lambda \omega_1 + (1-\lambda)\omega_2$ be a convex
combination of two states $\omega_1$ and $\omega_2$ in
$\boldsymbol{S}(M,\gb)$. Then $\omega_j \circ \alpha_{M,O} \in
\boldsymbol{F}(\pi_{\tilde{\omega}} \circ \alpha_{M,O})$, $j =1,2$, for some
state $\tilde{\omega} \in \tilde{\boldsymbol S}(M,\gb)$, and going
back to the definition of the folium, this shows in fact that $\omega'
\circ \alpha_{M,O} \in \boldsymbol{F}(\pi_{\tilde{\omega}} \circ
\alpha_{M,O})$. Thus $\omega' \in \boldsymbol{S}(M,\gb)$, showing that
$\boldsymbol{S}(M,\gb)$ is convex. {}\hfill $\Box$

Finally, we shall demonstrate that a 
locally normal and intermediate factorial 
state space induces a generally covariant realization of the principle
of local definiteness proposed by Haag, Narnhofer and Stein
\cite{HNS}. This principle was introduced in the context of a net of
observable algebras $\{\mathcal{A}(O)\}_{O\in\calK(M,\gb)}$ over a
fixed, globally hyperbolic background spacetime $(M,\gb)$. The
{\bf principle of local definiteness} demands
that there exists a Hilbert-space representation $\pi$ of the
$C^*$-algebra $\mathcal{A}$ generated by
$\{\mathcal{A}(O)\}_{O\in\calK(M,\gb)}$ so that the set of states,
$\mathcal{S}$, of the theory can be characterized as consisting of all
states $\omega$ on $\mathcal{A}$ that can be extended to normal states
on the local von Neumann algebras $\mathcal{M}(O) =
\pi(\mathcal{A}(O))''$, $O \in \calK(M,\gb)$. Furthermore, it was
required in \cite{HNS} that the local von Neumann algebras
$\mathcal{M}(O)$ are factors, at least for a suitable collection of
regions $O$. Here we take the point of view that one should replace
this condition by the (weaker) condition of intermediate factoriality
with respect to the family of local von Neumann algebras
$\{\mathcal{M}(O)\}_{O \in \calK(M,\gb)}$ since this avoids having to
specify precise geometric conditions on the regions $O$ for which
$\mathcal{M}(O)$ should be a factor.

Adopting this point of view, we may observe the following. Let $\euA$
be a locally covariant quantum field theory
with a locally normal and intermediate factorial state space
$\boldsymbol{S}$, and for $(M,\gb) \in \obj(\frakM)$, let
$\{\mathcal{A}(O)\}_{O\in\calK(M,\gb)}$ be the net of $C^*$-algebras on
$(M,\gb)$ induced by $\euA$ according Prop.\ 2.2. Let $\tilde{\omega}$
be any state in $\tilde{\boldsymbol{S}}(M,\gb)$ where
$\tilde{\boldsymbol{S}}$ is a locally quasi-equivalent state space to
which $\boldsymbol{S}$ is locally normal (cf.\ Def.\ 2.3(iii)), and
denote by $\tilde{\pi}$ the corresponding GNS-representation. This
representation induces a representation $\pi$ of $\euA$ via defining
the representations $\pi \rest \mathcal{A}(O)$ as $\tilde{\pi} \circ
\alpha_{M,O}^{-1}$, and hence it induces the corresponding net of von
Neumann algebras $\{\mathcal{M}(O)\}_{O\in\calK(M,\gb)}$. It is easy
to see that each state $\omega \in \boldsymbol{S}(M,\gb)$ extends to a
normal state on $\mathcal{M}(O)$ owing to local normality of
$\boldsymbol{S}$; additionally $\{\mathcal{M}(O)\}_{O\in\calK(M,\gb)}$
satisfies the condition of intermediate factoriality because
$\boldsymbol{S}$ is intermediate factorial. We formulate the result of
this discussion subsequently as
\begin{Pro} If $\boldsymbol{S}$ is locally normal and intermediate
  factorial, then
the set of states $\boldsymbol{S}(M,\gb)$ for
$\{\mathcal{A}(O)\}_{O\in\calK(M,\gb)}$ fulfills the principle of
local definiteness, for each $(M,\gb) \in \obj(M,\gb)$.
\end{Pro}
\subsection{State Space of the Klein-Gordon Field Distinguished by
  Microlocal Spectrum Condition} \label{3.2}
${}$\\[6pt]
For the locally covariant quantum field theory of the Klein-Gordon
field, we will show in the present subsection that
the microlocal spectrum condition selects a state space that is
locally quasi-equivalent and intermediate factorial.

 We have to provide some explanations first.
Let $(M,\gb) \in \obj(\frakM)$ and let $E$, $\mathfrak{W}({\mathcal
  R},\sigma)$ be defined with respect to the Klein-Gordon equatation
\eqref{KGeqn} on $(M,\gb)$.
A state $\omega$ on
$\mathfrak{W}({\mathcal R},\sigma)$ is called quasifree if its
two-point function
$$ w^{(\omega)}_2(f,h) = \left. \partial_t \partial_{\tau}\right|_{t =
  \tau = 0} \omega(W(tEf)W(\tau Eh))$$
exists for all $f,h \in C^{\infty}_0(M,\R)$, and if $\omega$ is
determined by $w^{(\omega)}_2$ according to
$$ \omega(W(Ef)) = {\rm e}^{-w^{(\omega)}_2(f,f)}\,.$$
A quasifree state $\omega$ is a Hadamard state if its two-point
function is of Hadamard form. This property is a constraint on the
short-distance behaviour of the two-point function.
Qualitatively, it means that $w_2^{(\omega)}$ is a distribution on
$C^{\infty}_0(M,\R)\times C_0^{\infty}(M,\R)$ of the form
\begin{equation} \label{HadForm}
 w_2^{(\omega)}(f,h) = \lim_{\epsilon\to 0}
\,\int(G_{\epsilon}(x,y) + H_{\omega}(x,y))
f(x)h(y)\,d\mu_{\gb}(x)\,d\mu_{\gb}(y)
\end{equation}
where $H_{\omega}$ is a smooth integral kernel depending on the state
$\omega$, while the singular part of $w^{(\omega)}_2$ is given as the
limit of a family of integral kernels $G_{\epsilon}$ which are
determined by the metric $\gb$ and the Klein-Gordon equation via the
so-called Hadamard recursion relations. The leading singularity is of
the type 1/(squared geodesic distance from $x$ to $y$). We refer to
 \cite{KW} for details. The Hadamard property 
can be equivalently expressed in terms of a condition on
the wavefront set ${\rm WF}(w^{(\omega)}_2)$ of the two-point function
\cite{Rad1} (see also \cite{SV2}): $\omega$ is a Hadamard state
exactly if the pairs of covectors $(x,\eta)$ and $(x',\eta')$ which
are in ${\rm WF}(w^{(\omega)}_2)$ are such that their base-points $x$ and
$x'$ lie on a lightlike 
geodesic, and the co-tangent vectors $\eta$ and $-\eta'$
are co-tangent and co-parallel to that geodesic, with $\eta$
future-pointing. 

This characterization of the Hadamard condition in terms of a
constraint on the two-point function of a state is also referred to a
``microlocal spectrum condition'' because it mimicks the usual, flat
space spectrum condition in the sense of microlocal analysis; its
advantage is that it may be formulated for general quantum field
theories, in contrast to the Hadamard condition which requires that
the $2$-point function satisfies a hyperbolic wave-equation \cite{BFK,Ver4}.
We refer to the indicated references for further discussion. In the
context of the present subsection, we will use ``Hadamard condition'' and
``microlocal spectrum condition'' synonymously.

Now let $\euA$ be the locally covariant quantum field theory
associated with the Klei-Gordon field as in Subsec.\ \ref{kgfe}.
It is important to note that,
owing to the functorial transformation properties of wavefront
sets under diffeomorphisms \cite{Horm}, a quasifree Hadamard state
$\omega'$ on $\euA(M',\gb')$ induces a quasifree Hadamard state
$\omega' \circ \alpha_{\psi}$ on $\euA(M,g)$ whenever $\psi
\in \Hom_{\frakM}((M,\gb),(M',\gb'))$. 
Furthermore,
was shown in \cite{FNW} that there exists a large set of quasifree
Hadamard states for the Klein-Gordon field on every globally hyperbolic
spacetime $(M,\gb)$. Moreover, the results in \cite{Ver1} show
that the GNS-representations of quasifree Hadamard states are locally
quasi-equivalent, and in \cite{Ver3} it was proved that the condition
of intermediate factoriality is fulfilled for quasifree Hadamard states. 
We may thus summarize these results in the subsequent
\begin{Thm}
For each $(M,\gb) \in \obj(\frakM)$, define $\boldsymbol{S}(M,\gb)$ as
the set of all states  
on $\euA(M,\gb)$ whose GNS-representations are locally
quasiequivalent to the GNS-representation of 
any quasifree Hadamard state on $\euA(M,\gb)$. This
assignment results in a state space which is locally quasi-equivalent
and intermediate factorial, and $\boldsymbol{S}(M,\gb)$ contains in
particular all quasifree Hadamard states on $\euA(M,\gb)$.
\end{Thm}
\section{Dynamics} \label{Dynamics}
\subsection{Relative Cauchy-Evolution}
${}$\\[6pt]
For theories obeying the time-slice axiom one 
can define relative Cauchy-evo\-lu\-tions, as follows. Let $(M_1,\gb_1)$
and $(M_2,\gb_2)$ be in $\obj(\frakM)$. We suppose that there are
globally hyperbolic sub-regions $N^{\pm}_j$ of $M_j$, $j=1,2$ containing
Cauchy-surfaces of the respective spacetimes. Moreover, we assume that
there are isometric (and
orien\-tation/time-orien\-tation-pre\-ser\-ving) diffeomorphisms
$\iota^{\pm} : N^{\pm}_1 \to N_2^{\pm}$ when the regions are endowed
with the appropriate restrictions of the metrics $\gb_1$ and $\gb_2$,
respectively. Henceforth, we shall suppress the diffeomorphisms
$\iota^{\pm}$ in our notation and identify $N_1^{\pm}$ and $N_2^{\pm}$
as being equal. The isometric embeddings of $N_j^{\pm}$ into $M_j$
will be denoted by $\psi^{\pm}_j$. They are depicted in the following
diagram:
\begin{equation*}
\begin{CD}
N_1^+ @>{\psi^+_1}>> M_1 
@<{\psi^-_1}<<N_1^-\\
@| @. @| \\
N_2^+ @>{\psi^+_2}>> 
M_2 @<{\psi^-_2}<<N_2^-
\end{CD}
\end{equation*}
By the functorial properties of a locally covariant 
quantum field theory $\euA$, the previous diagram gives rise to
the next:
\begin{equation*}
\begin{CD}
\euA(N_1^+) @>{\alpha_{\psi^+_1}}>> \euA(M_1) 
@<{\alpha_{\psi^-_1}}<<\euA(N_1^-)\\
@| @. @| \\
\euA(N_2^+) @>{\alpha_{\psi^+_2}}>> 
\euA(M_2) @<{\alpha_{\psi^-_2}}<<\euA(N_2^-)
\end{CD}
\end{equation*}
where we have, for the sake of simplicity, suppressed the appearence
of the spacetime metrics in our notation. If the theory $\euA$ obeys
the time-slice axiom, then all the morphisms in this diagram are onto
and invertible, and hence one obtains from it an automorphism 
$\beta \in \Hom_{\frakA}(\euA(M_1),\euA(M_1))$ by setting
$$ \beta = \alpha_{\psi^-_1}\circ \alpha_{\psi^-_2}^{-1} \circ
\alpha_{\psi_2^+} \circ \alpha_{\psi_1^+}^{-1}\,.$$
Under certain circumstances (which may be expected to be generically
fulfilled)  it is possible to form the functional derivative
of the relative Cauchy-evolution with respect to the metrics of the
spacetimes involved in its construction. This functional derivative
then has the meaning of an energy-momentum tensor. In fact, we will
show below for the example of the Klein-Gordon field that the
functional derivative of the relative Cauchy-evolution agrees with the
action of the quantized energy-momentum tensor in representations of
quasifree Hadamard states.

In order to give these ideas a more precise shape, we introduce the
following
\\[6pt]
{\bf Geometric assumptions.}
\begin{itemize}
\item We consider a globally hyperbolic spacetime $(M,\gbo)$ where it
  is assumed that $M$ can be covered by a single coordinate patch.
\item We pick a Cauchy-surface $C$ in $(M,\gbo)$, and two open
  subregions $N_{\pm}$ of $M$ with the properties:
\begin{itemize}
\item $\overline{N_{\pm}} \subset {\rm int}\, J^{\pm}(C)$,
\item $(N_{\pm},\gbo_{N_{\pm}})$ are contained in $\obj(\frakM)$,
\item $N_{\pm}$ contain
  Cauchy-surfaces for $(M,\gbo)$.
\end{itemize}
\item Let $G$ be a set of Lorentzian metrics on $M$ with the following
  properties: 
\begin{itemize}
\item Each $\gb \in G$ deviates from $\gbo$ only on a compact subset
  of the region
$$ M_{(+,-)} = M \backslash {\rm cl}[J^-(N_-) \cup J^+(N_+)]\,, $$
\item each $(M,\gb)$, $\gb \in G$, is a member of $\obj(\frakM)$,
\item $C$ is a Cauchy-surface for $(M,\gb)$, $\gb \in G$,
\item  If $\gb \in G$, then the family $\{\gb_s\}_{s \in [0,1]}$
  of metrics given by
$$ \gb_s = \gbo + s(\gb -\gbo)\,, \quad 1 \ge s \ge 0\,,$$
is also contained in $G$.
\item If $\phi$ is a diffeomorphism of $M$ which acts trivially
  outside of $M_{(+,-)}$, then $\phi_*\gb$ is contained in $G$.
\end{itemize}
\end{itemize}
It is clear from these assumptions that one may view the metrics $\gb$
in $G$ as ``perturbations'' around the metric $\gbo$ on
$M_{(+,-)}$. Moreover, $(N_{\pm},\gbo_{N_{\pm}})$ are also globally
hyperbolic submanifolds of $(M,\gb)$ for each $\gb\in G$. Hence there
are isometric embeddings $\psi^{\pm}_{\gb} \in
\Hom_{\frakM}((N_{\pm},\gbo_{N_{\pm}}),(M,\gb))$ for all $\gb \in G$
as well as isometric embeddings $\psi_{\circ}^{\pm} \in
\Hom_{\frakM}((N_{\pm},\gbo_{N_{\pm}}),(M,\gbo))$. To these embeddings
one can associate the relative Cauchy-evolution $\beta_{\gb} \in
\Hom_{\frakA}(\euA(M,\gbo),\euA(M,\gbo))$ given by
\begin{equation}
\label{cauchyev}
 \beta_{\gb} = \alpha_{\psi_{\circ}^-} \circ \alpha^{-1}_{\psi_{\gb}^-}
 \circ \alpha_{\psi_{\gb}^+}\circ  \alpha_{\psi_{\circ}^+}^{-1}\,.
\end{equation}
{\it Remarks. } 
(A) One may view $\beta_{\gb}$ as a ``scattering morphism'' describing
the change that the propagation of a quantum field undergoes passing
through the region with the ``metric perturbation'' $\gb-\gbo$
compared to the background metric $\gbo$.
\\[6pt]
(B)
There is some relation between the relative Cauchy-evolution and the
evolution of Cauchy-data from one Cauchy-surface to another which
e.g.\ in the case of the scalar Klein-Gordon field is also known to
lead to $C^*$-algebraic endomorphisms \cite{KayCE,TorVar}. We refer to the
references for more discussion. 
\\[6pt] 
(C)
Hollands and Wald \cite{HW} consider for the case of
the free Klein-Gordon field related operators $\tau_{\gb}^{\rm adv}$
and $\tau_{\gb}^{\rm ret}$, which would correspond to the operators
$\alpha_{\psi^+_{\circ}} \circ \alpha_{\psi_{\gb}^+}^{-1}$ and 
$\alpha_{\psi^-_{\circ}} \circ \alpha_{\psi_{\gb}^-}^{-1}$.
\\[10pt]
As the theory $\euA$ is locally covariant, it follows that
the relative Cauchy-evolution is insensitive to changing $\gb$ into
$\phi_*\gb$ when $\phi$ is a diffeomorphism of $M$ that acts trivially
outside of the intermediate region $M_{(+,-)}$. More precisely, one
obtains:
\begin{Pro} \label{Diffeoinv}
Let $\phi$ be a diffeomorphism of $M$ that acts trivially outside of
$M_{(+,-)}$ (i.e.\ $\phi(x) =x$ for all $x$ in the complement of
$M_{(+,-)}$).
Then 
 $$ \beta_{\gb} = \beta_{\phi_*\gb} \,, \quad \gb \in G\,.$$
\end{Pro}
\noindent
{\it Proof.} It holds that $\phi$ is a morphism in
$\Hom_{\frakM}((M,\gb),(M,\phi_*\gb))$, and  hence
$$ \phi \circ \psi^{\pm}_{\gb} = \psi^{\pm}_{\phi_*\gb} $$
owing to the definition of $\psi^{\pm}_{\gb}$ since $\phi$ acts
trivially on $N_{\pm}$. On the other hand, it
holds that
\begin{eqnarray*}
\beta_{\gb}& =& \alpha_{\psi_{\circ}^-} \circ \alpha^{-1}_{\psi_{\gb}^-}
 \circ \alpha_{\psi_{\gb}^+}\circ  \alpha_{\psi_{\circ}^+}^{-1}\\
&=& \alpha_{\psi_{\circ}^-} \circ \alpha^{-1}_{\psi_{\gb}^-} \circ
\alpha_{\phi}^{-1} \circ \alpha_{\phi}
 \circ \alpha_{\psi_{\gb}^+}\circ  \alpha_{\psi_{\circ}^+}^{-1}\\
&=& \alpha_{\psi_{\circ}^-} \circ \alpha^{-1}_{\phi \circ \psi_{\gb}^-}
 \circ \alpha_{\phi \circ\psi_{\gb}^+}\circ  \alpha_{\psi_{\circ}^+}^{-1}\\
&=&\alpha_{\psi_{\circ}^-} \circ \alpha^{-1}_{\psi_{\phi_*\gb}^-}
 \circ \alpha_{\psi_{\phi_*\gb}^+}\circ  \alpha_{\psi_{\circ}^+}^{-1}\\
&=&  \beta_{\phi_*\gb}\,.
\end{eqnarray*}
${}$ \hfill $\Box$
${}$\\[6pt]
We will now make assumptions that allow us to define the functional
derivative of $\beta_{\gb}$ with respect to $\gb \in G$. To this end,
we assume that $\pi$ is a Hilbert-space representation of $\euA
(M,\gbo)$, and that there is a dense subspace $\mathcal{V}$ of the
representation-Hilbert-space $\mathcal{H}$ and a
dense $*$-sub-algebra $\mathcal{B}$ of $\euA (M,\gbo)$ so that, for all
smooth curves $[0,1] \owns s \mapsto \gb_s \in G$ with $\gb_{(s=0)} =
\gbo$, there holds
$$ \left.\frac{d}{ds} \langle \theta, \pi(\beta_{\gb_s}(B))\theta
  \rangle \right|_{s=0} = \int_M b^{\mu \nu}(x) \delta\gb_{\mu\nu}(x)
(-\ggo(x))^{1/2}\,dx $$
for all $\theta \in \mathcal{V}$, $B \in \mathcal{B}$ with a suitable
smooth section $x \mapsto b^{\mu\nu}(x)$ in $TM \otimes TM$
 (depending on $\theta$ and $B$);
 we have written $\delta \gb
= \left. d\gb_s/ds\right|_{s=0}$,
 and $\ggo$ is the determinant of $\gbo$ in the
coordinates used for $M$.
Then we write
$$ \langle \theta ,\frac{\delta}{\delta \gb_{\mu\nu}(x)}\pi(\beta_{\gb}B)
      \theta \rangle = b^{\mu\nu}(x)\,, $$
and thus the functional derivative of the relative
Cauchy-evolution $\beta_{\gb}$ with
respect to the metric $\gb$,
$$ \frac{\delta}{\delta \gb_{\mu\nu}(x)}\pi(\beta_{\gb}B)\,, $$
is defined in the representation $\pi$ for all $B \in \mathcal{B}$ in
the sense of quadratic forms on $\mathcal{V}$. 
(As announced before, these assumptions are realized for the free
scalar Klein-Gordon field in representations of quasifree Hadamard
states, see Sec.\ \ref{ReCEKG} below). The functional derivative of
$\beta_{\gb}$ with respect to $\gb$ describes the reaction of the
quantum system to an infinitesimal local change of the spacetime
metric. As known in classical field theory, this is described by the
energy-momentum tensor, and we will find this corroborated in the
quantum field case by Thm \ref{funderiv} below. It is mentioned in
\cite{HW} that the functional derivative of $\tau^{\rm adv/ret}_{\gb}$
with respect to $\gb$ describes the advanced/retarded response of the
quantum system upon infinitesimal metric changes.
\par
When the indicated assumptions are fulfilled, then we find that the
relative Cauchy-evolution is divergence-free.
\begin{Thm} \label{Divergencefree}
For all $B \in \mathcal{B}$, one has
$$ \nabla_{\mu} \frac{\delta}{\delta
  \gb_{\mu\nu}(x)}\pi(\beta_{\gb}(B)) = 0\,, \quad x \in M\,, $$
in the sense of quadratic forms on $\mathcal{V}$ where $\nabla$
is the covariant derivative with respect to $\gbo$.
\end{Thm}
\noindent
{\it Proof.} Let $X$ be a smooth vector field on $M$ which vanishes
outside of $M_{(+,-)}$, and let $\phi_s$, $s \in \R$, be the
one-parametric group of diffeomorphisms that is generated by $X$.
By Prop.\ \ref{Diffeoinv}, we have $\beta_{\gbo} -
\beta_{\phi_{s*}\gbo} = 0$ for all $s$, and hence one obtains that
$$ \frac{d}{ds}\beta_{\phi_{s*}\gbo} = 0\,.$$
On the other hand, using the
notation $b^{\mu\nu}(x) = \langle \theta,\delta
\pi(\beta_{\gb}(B))/\delta\gb_{\mu\nu}(x)\theta\rangle$ 
and recalling the definition of $\delta
\beta_{\gb}/\delta \gb_{\mu\nu}(x)$, we have
$$ 0 = \left. \frac{d}{ds}\langle \theta,
  \pi(\beta_{\phi_{s*}\gbo}(B))\theta \rangle\right|_{s = 0} =
  \int_M b^{\mu\nu}(x)\left.\frac{d}{ds}\right|_{s=0}\phi_{s*}\gbo_{\mu\nu}(x)
(-\ggo)^{1/2}(x)\,dx  
  $$
for all $B \in \mathcal{B}$, $\theta \in \mathcal{V}$. 
Now one can conclude that $\nabla_{\mu}b^{\mu\nu} = 0$ as in the case
of classical field theory (cf.\ \cite{HE}, Sec.\ 3.3): It holds that
$\left.\frac{d}{ds}\right|_{s=0}\phi_{s*}\gbo_{\mu\nu} = \pounds_X
\gbo_{\mu\nu}= \nabla_{\mu}X_{\nu} + \nabla_{\nu}X_{\mu}$, where
  $\pounds_X$ denotes the Lie-derivative, and hence
\begin{eqnarray*}
 \lefteqn{0 = \int_M
   b^{\mu\nu}(x)\pounds_X\gbo_{\mu\nu}(x)(-\ggo)^{1/2}(x)\,
dx}\\
& = & 2\int_M (\nabla_{\mu}(b^{\mu\nu}X_{\nu})(x) -
(\nabla_{\mu}b^{\mu\nu}(x))X_{\nu}(x)) \,(-\ggo)^{1/2}(x)\,dx\,.
\end{eqnarray*}
The first term in the last expression is a divergence and can be
converted to a surface integral  which hence vanishes since $X$ has
compact support. As $X$ was an arbitrary vectorfield supported inside
$M_{(+,-)}$, one thus concludes that $\nabla_{\mu}b^{\mu\nu}(x)= 0$
for $x \in M_{(+,-)}$; on the other hand, $b^{\mu\nu}(x) = 0$ for all
$x$ outside of $M_{(+,-)}$ according to the definition of the
functional derivative of the Cauchy-evolution. Thus
$\nabla_{\mu}b^{\mu\nu} =0$ on $M$, and this completes the proof.
{} \hfill $\Box$  
\subsection{Relative Cauchy-Evolution for the Klein-Gordon Field}
\label{ReCEKG} 
${}$\\[6pt]
In the present sub-section we will investigate the relation between
the functional derivative of the relative Cauchy-evolution for the
quantum Klein-Gordon field with respect to the spacetime metric, and
the quantum field's energy-momentum tensor. This will be the presented
in Theorem \ref{funderiv} below.
 Before stating this result, we will discuss the form of
the relative Cauchy-evolution for the generally covariant Klein-Gordon
field in some detail.

Let $(M,\gb)$ be an object in $\obj(\frakM)$ and let $(N,\gb_N)$ be
globally hyperbolic sub-spacetime of $(M,\gb)$, so that the identical
injection $\iota_N: N \to M$, $\iota_N(x) = x$ is a morphism in
$\Hom_{\frakM}((N,\gb_N),(M,\gb))$, where $\gb_N$ is $\gb$ restricted
to $N$. Furthermore, let $(\mathcal{R},\sigma)$ denote the symplectic
space of solutions of the Klein-Gordon equation \eqref{KGeqn} on $(M,\gb)$,
and $(\mathcal{R}_N,\sigma_N)$ the corresponding symplectic space of
solutions on $(N,\gb_N)$. $E$ and $E_N$ will denote the associated
propagators, respectively. We have seen above that $\iota_N$ induces a
$C^*$-endomorphism $\alpha_{\iota_N}:
\mathfrak{W}(\mathcal{R}_N,\sigma_N) \to
\mathfrak{W}(\mathcal{R},\sigma)$ by
$$ \alpha_{\iota_N} (W_N(\varphi)) = W(T_N\varphi)\,, \quad \varphi \in
\mathcal{R}_N\,,$$
where we have denoted by $W_N(\,.\,)$ the Weyl-generators of
$\mathfrak{W}(\mathcal{R}_N,\sigma_N)$ and by $W(\,.\,)$ those of
$(\mathcal{R},\sigma)$. The map $T_N$ assigns to each element $E_Nf$,
$f\in C^{\infty}_0(N,\R)$, of $\mathcal{R}_N$ the element $Ef \in
\mathcal{R}$. 

Let us now consider the case where $N$ contains a Cauchy-surface for
$(M,\gb)$. In this case, Dimock \cite{Dim.KG} has shown that the map
$T_N$ is surjective, i.e.\ $T_N\mathcal{R}_N = \mathcal{R}$. $T_N$ is
also injective (since it is symplectic), and we want to derive the
form of the inverse map $T_N^{-1}$. To this end, let $\varphi \in
\mathcal{R}$, and let $\Sigma$ be a Cauchy-surface for $(M,\gb)$
contained in $N$. There exists a pair of two other Cauchy-surfaces
$\Sigma^{\rm adv}$ and $\Sigma^{\rm ret}$ for $(M,\gb)$ in $N$ where 
$\Sigma^{\rm adv}$ lies
in the timelike future and $\Sigma^{\rm ret}$ in the timelike past of
$\Sigma$, hence $U = {\rm int}\, J^-(\Sigma^{\rm adv}) \cap
J^+(\Sigma^{\rm ret})$ is
an open neighbourhood of $\Sigma$ whose closure in contained in
$N$. Now we choose a partition of unity $\{\chi^{\rm adv},\chi^{\rm
  ret}\}$ of $M$ so that $\chi^{\rm adv} = 0$ on $J^-(\Sigma^{\rm ret})$ and
$\chi^{\rm ret} = 0$ on $J^+(\Sigma^{\rm adv})$. Then the properties
$\chi^{\rm adv} + \chi^{\rm ret} = 1$ and $(\nabla^a\nabla_a +
 \xi R + m^2) \varphi = 0$ imply
\begin{equation} \label{KGref}
(\nabla^a\nabla_a +
\xi R + m^2) (\chi^{\rm adv}\varphi) = - (\nabla^a\nabla_a +
 \xi R + m^2) (\chi^{\rm ret}\varphi)\,.
\end{equation}
Since the left-hand side vanishes on $J^-(\Sigma^{\rm ret})$ and the
right-hand side vanishes on $J^+(\Sigma^{\rm adv})$ while $\varphi = E f$ has
support in $J({\rm supp}\,f)$ for some compactly supported $f$, one
deduces that both the left- and right-hand side expressions of
\eqref{KGref} are compactly supported in $\overline{U} \subset
N$. Using the properties of the propagator $E$, one can moreover show
(cf.\ \cite{Dim.KG})
$$ E(\nabla^a\nabla_a +
 \xi R + m^2) (\chi^{\rm adv/ret}\varphi) = \pm \varphi\,, \quad
\varphi \in \mathcal{R}\,.$$
Since $E(\nabla^{\mu}\nabla_{\mu} +
m^2 + \xi R) (\chi^{\rm adv/ret}\varphi)$ is contained in
$E(C^{\infty}_0(N,\R))$ and $Ef \mapsto E_Nf$, $f \in
C_0^{\infty}(N,\R)$, is a symplectic map from $(\mathcal{R},\sigma)$
onto $(\mathcal{R}_N,\sigma_N)$ owing to the uniqueness of advanced
and retarded fundamental solutions of the Klein-Gordon equation in
globally hyperbolic spacetimes, we can see that $T_N^{-1}:
(\mathcal{R},\sigma) \to (\mathcal{R}_N,\sigma_N)$ is given by
$$ T^{-1}_N(\varphi) = \pm E_N(\nabla^a\nabla_a +
 \xi R + m^2)(\chi^{\rm adv/ret} \varphi)\,.$$
\par
Now we wish to study the relative Cauchy-evolution for the scalar
Klein-Gordon field. We assume that we are in the situation 
described in the previous subsection: We are
given a globally spacetime $(M,\gbo)$, with subregions $N_{\pm}$ and
$M_{(+,-)}$ on the latter of which metrics $\gb$ in a set $G$ deviate
from $\gbo$,
where these data are subject to the geometric assumptions listed above.
\par
 For the generally covariant theory of the Klein-Gordon field, we see
 from our discussion above that $\beta_{\gb}$ acts on the generators
 $\Wo(\,.\,)$ of the CCR-algebra of the Klein-Gordon field on
 $(M,\gbo)$ like
$$ \beta_{\gb}(\Wo(\varphi)) = \Wo(F_{\gb}\varphi)\,;$$
here, $F_{\gb}: \calRo \to \calRo$ is the symplectic map
$$ F_{\gb} = T_{N_-,\nnn} \circ T_{N_-,\gb}^{-1} \circ T_{N_+,\gb}
\circ T_{N_+,\nnn}^{-1}$$
with
\begin{eqnarray*}
 & T_{N_{\pm},\gb} : E_{N_{\pm},\gb}f \mapsto
 E_{\gb}\iota_{N_{\pm}}{}_*f\,, \quad f \in
 C_0^{\infty}(N_{\pm},\R)\,,\\
& T_{N_{\pm},\nnn} : \Eo_{N_{\pm}} f \mapsto \Eo
\iota_{N_{\pm}}{}_*f\,, \quad f \in C_0^{\infty}(N_{\pm},\R)\,,\\
& T_{N_{\pm},\gb}^{-1}: \phi \mapsto -
E_{N_{\pm},\gb}K_{\gb}(\chi_{\pm}^{\rm ret}\phi)\,, \quad
\phi \in \mathcal{R}_{\gb}\,,\\
& T_{N_{\pm},\nnn}^{-1} : \varphi \mapsto -
\Eo_{N_{\pm}}\Ko(\chi_{\pm}^{\rm ret}\varphi)\,, \quad \varphi \in
\calRo\,,
\end{eqnarray*}
where $\Eo,\calRo,\sigmao$,
$\Eo_{N_{\pm}},\calRo_{N_{\pm}},\sigmao_{N_{\pm}}$,
$E_{\gb},\mathcal{R}_{\gb},\sigma_{\gb}$ and
$E_{N_{\pm},\gb},\mathcal{R}_{N_{\pm},\gb},\sigma_{N_{\pm},\gb}$
denote the propagators, range-spaces and symplectic forms
corresponding to the Klein-Gordon equation on the spacetimes
$(M,\gbo)$, $(N_{\pm},\gbo_{N_{\pm}})$, $(M,\gb)$ and $(M,\gbo)$,
respectively.
The functions $\chi^{\rm adv/ret}_{\pm}$ are defined relative to
suitable pairs of Cauchy-surfaces $\Sigma^{\rm adv/ret}_{\pm}$ 
lying in $N_{\pm}$. By $K_{\gb}$ and $\Ko$
we denote the Klein-Gordon operator
$$ \nabla^a\nabla_a + \xi R + m^2$$
on the spacetimes $(M,\gb)$ and $(M,\gbo)$, respectively. Note that
(up to identification) $E_{N_{\pm},\gb} = \Eo_{N_{\pm}}$ for all $\gb
\in G$ according to our geometric assumptions, and thus also
$\mathcal{R}_{N_{\pm},\gb}=\calRo_{N_{\pm}}$, $\sigma_{N_{\pm},\gb} =
\sigmao_{N_{\pm}}$. This entails
\begin{equation}
\label{formulauno}
F_{\gb}\varphi =  \Eo K_{\gb}\chi_-^{\rm ret}E_{\gb}\Ko\chi_+^{\rm
  ret}\varphi\,, \quad \varphi \in \calRo\,,
\end{equation}
where we have dropped the embedding identifications $\iota_{N_{\pm}}{}_*$
from our notation. This relation will be the key ingredient in the
proof of the next theorem. Prior to stating it, some further
preparation is required.

Let us select some arbitrary quasifree Hadamard state $\omega$ on
$\euA(M,\gbo)= \mathfrak{W}(\calRo,\sigmao)$, the Weyl-algebra of the
Klein-Gordon field on $(M,\gbo)$. Then we will write
$$ W_{\omega}(\varphi) = \pi_{\omega}(\Wo(\varphi))\,,\quad \varphi \in
\calRo\,,$$
for the Weyl-operators in the GNS-representation $\pi_{\omega}$ of
$\omega$; then we have 
$$ W_{\omega}(\varphi) = {\rm e}^{i\Fco(\varphi)} $$
with suitable selfadjoint operators $\Fco(\varphi)$ in the
GNS-Hilbert-space $\mathcal{H}_{\omega}$, depending linearly on
$\varphi$, and 
$$ w_2^{(\omega)}(f,h) = \langle \Omega_{\omega},\Fco(\Eo
f)\Fco(\Eo h)\Omega_{\omega} \rangle \,, \quad f,h \in
C_0^{\infty}(M,\R)\,,$$
with the GNS-vector $\Omega_{\omega}$. Let $\mathcal{V}_{\omega}$ the
set of all vectors $\theta$ in $\mathcal{H}_{\omega}$ which are of the
form $\theta = B\Omega_{\omega}$ where $B$ is an arbitrary polynomial in
the variables $W_{\omega}(\varphi),\Fco(\varphi')$ as
$\varphi$ and $\varphi'$ vary over $\calRo$. One can show that each
$\theta \in \mathcal{V}_{\omega}$ is in the domain of all operators
$\Phi_{\omega}(\varphi)$ and that the wavefront sets WF$(w^{[\theta]}_2)$
of the two-point functions induced by $\theta \in \mathcal{V}_{\omega}$,
$$ w^{[\theta]}_2(f,h) = \langle \theta, \Fco(\Eo
f)\Fco(\Eo h)\theta \rangle\,, \quad f,h \in
C_0^{\infty}(M,\R)\,,$$
are of the same shape as those of the two-point functions of Hadamard
states \cite{FewVer2}. Furthermore, denoting
by
$$ \Fo(f) = \Fco(Ef)\,,\quad f \in C_0^{\infty}(M,\R)\,,$$
the quantum field induced by $\Fco$, one can show that there is for
each pair of vectors $\theta,\theta' \in {\mathcal V}_{\omega}$ a
smooth function $x \mapsto \langle \theta,\Fo(x)\theta'\rangle$ on $M$
so that
$$ \langle \theta,\Fo(f)\theta'\rangle
 = \int_M \langle \theta,\Fo(x)\theta'\rangle f(x)\,
(-\ggo)^{1/2}(x)\,dx $$
where we recall that 
$\ggo(x)$ is the determinant of $\gbo$ in the coordinates used
for $M$. These assertions rest on the fact that (1) $\Omega_{\omega}$
is an analytic vector for all $\Fco(\varphi)$, (2)
$[\Fco(\varphi),W_{\omega}(\tilde{\varphi})] = -
\sigma(\varphi,\tilde{\varphi})W_{\omega}(\tilde{\varphi})$, and
iterated use of this relation, (3) the distribution $f \mapsto
w^{(\omega)}_2(f,h)$ is induced by a smooth function, and
$w^{[\theta]}_2(f,h)$ can be 
reduced to a sum of products of such $w^{(\omega)}_2(f,h_j)$ (with
suitable coefficients) since $\omega$ is quasifree.

After these preparations, we obtain:
\begin{Thm} \label{funderiv}
Under the geometric assumptions listed above, there holds 
\begin{equation} \label{derivative}
 \frac{\delta}{\delta
  \gb_{\mu\nu}(x)}(\beta_{\gb}\Wo)_{\omega}(\varphi) =
-\frac{i}{2}[T^{\mu\nu}(x),W_{\omega}(\varphi)]\,, \quad \varphi \in
\calRo,\ x \in M_{(+,-)}\,,
\end{equation}
in the sense of quadratic forms on $\mathcal{V}_{\omega}$, where
$T_{\mu\nu}$ is the generally covariant 
renormalized energy-momentum tensor of the quantized
Klein-Gordon field on $(M,\gbo)$ in the GNS-repre\-sen\-ta\-tion 
of $\omega$, and
$\omega$ is an arbitrary quasifree Hadamard state.
\end{Thm}
\noindent 
{\it Remarks.} (A) Note that the classical expression for $T_{\mu\nu}$
is $T_{\mu\nu}= \left.\frac{2}{\sqrt{-g}}\frac{\delta}{\delta
  \gb^{\mu\nu}}S_{\rm KG}\right|_{\gb = \gbo}$ 
where $S_{\rm KG}$ is the action integral
of the Lagrangian density
$$ \mathcal{L}_{\rm KG}=
\frac{1}{2}\sqrt{-g}\left(\gb^{\mu\nu}\nabla_\mu\varphi\nabla_{\nu}\varphi
- (m^2 + \xi R)\varphi^2\right)\,.$$
Here we use the convention that $T_{\m\nu}$ is defined in this way,
and that $T^{\mu\nu} =
\gbo{}^{\mu\alpha}\gbo{}^{\nu\beta}T_{\alpha\beta}$ and {\it not}
$T^{\mu\nu}= \left.\frac{2}{\sqrt{-g}}\frac{\delta}{\delta
  \gb_{\mu\nu}}S_{\rm KG}\right|_{\gb = \gbo}$. The latter expression
differs from the former, which we use, by a sign.
\\[6pt]
(B) Instead of the generally covariant renormalized energy-momentum tensor one
may also use the energy-momentum tensor renormalized with respect to
$\omega$ as reference state, since the two definitions differ by a
term which is a multiple of the unit operator and hence is cancelled by
the commutator on the right hand side of \eqref{derivative}. In fact,
one may even use (after point-split regularization) the
``unrenormalized, formal  
expression'' (cf.\ \cite{Wald77}) for the quantum energy-momentum
tensor since only the commutator of the
energy-momentum tensor appears.
\\[6pt]
(C) Similarly one can show that
$$ \frac{\delta}{\delta\gb_{\mu\nu}(x)}P_{\gb} = -\frac{i}{2}[T^{\mu\nu}(x),P]
$$
holds in the sense of quadratic forms on $\mathcal{V}_{\omega}$ for
all polynomials
$$ P = \sum_{j \le \ell,\,k_j \le n}\Fco(\varphi_{j,1}) \cdots
\Fco(\varphi_{j,k_j})$$
in the field operators, with  
$$ P_{\gb} = \sum_{j \le \ell,\,k_j \le n}\Fco(F_{\gb}\varphi_{j,1}) \cdots
\Fco(F_{\gb}\varphi_{j,k_j})\,.$$
\\[10pt]
{\it Proof.} We will give the proof only for the case $\xi = 0$ in
order to simplify notation; however, the case of arbitrary $\xi$ can
be carried out along the same lines. 
For any smooth curve $[0,1] \owns s \mapsto \gb_s \in G$ with
$\gb_{s=0} = \gbo$, we will write $\delta\gb=
\left.d\gb_s/ds\right|_{s=0}$,
 and $\delta y_{\gb} = \left. \frac{d}{ds}\right|_{s=0}y_{\gb_s}$ 
for any function $y_{\gb}$ depending on $\gb \in G$. 

Let $\theta \in \mathcal{V}_{\omega}$. Since
$\beta_{\gb}(\Wo(\varphi)) = \Wo(F_{\gb}\varphi)$, one finds by a
general argument (cf.\ e.g.\ \cite{FewVer2}) that
$$ \delta(\beta_{\gb}W)_{\omega}\theta =
\delta(W(F_{\gb}\varphi))_{\omega}\theta =
\frac{i}{2}\{\Fco(\delta F_{\gb}\varphi),W_{\omega}(\varphi)\}\theta\,, \quad
\varphi \in \calRo\,,$$
where $\{A,B\} = AB + BA$ denotes the anti-commutator. One must
therefore derive an expression for $\delta F_{\gb}\varphi$. It holds
that (cf.\ \eqref{formulauno})
\begin{eqnarray*}
\delta F_{\gb}\varphi & = & \delta (\Eo K_{\gb}\chi^{\rm
  ret}_-E_{\gb}\Ko\chi^{\rm ret}_+ \varphi)\\
& = & \Eo(\delta K_{\gb})\chi^{\rm ret}_- \varphi + \Eo \Ko\chi^{\rm
  ret}_-(\delta E_{\gb})\chi^{\rm ret}_+\varphi\,.
\end{eqnarray*}
Now $\delta K_{\gb}$ is a partial differential operator whose
coefficient functions are compactly supported within $M_{(+,-)}$ as a
consequence of the geometric assumptions. Since $M_{(+,-)} \cap
J^-(N_-) = \emptyset$, and ${\rm supp}\,\chi^{\rm ret}_- \subset
J^-(N_-)$, it follows that $\Eo(\delta K_{\gb}\chi^{\rm ret}_-)\varphi
= 0$, and hence
$$ \delta F_{\gb}\varphi = \Eo \Ko \chi^{\rm ret}_-(\delta
E_{\gb})\Ko\chi^{\rm ret}_+\varphi\,.$$
On the other hand, it holds that
$$ \chi^{\rm ret}_- E_{\gb}\Ko \chi^{\rm ret}_+\varphi = 
 \chi^{\rm ret}_-E^{\rm adv}_{\gb}\Ko\chi^{\rm ret}_+\varphi -
 \chi^{\rm ret}_-E^{\rm ret}_{\gb}\Ko \chi^{\rm ret}_+\varphi\,,$$
and since $E^{\rm adv}_{\gb}\Ko\chi^{\rm ret}_+$ has support in
$J^+(N_+)$, while $\chi^{\rm ret}_-$ has support in $J^-(N_-)$, the
first term on the right hand side vanishes, leaving us with
$$ \delta F_{\gb}\varphi = - \Eo \Ko \chi^{\rm ret}_- (\delta
E_{\gb}^{\rm ret})
\Ko\chi^{\rm ret}_+\varphi\,.$$
Then we deduce from $E_{\gb}^{\rm ret}K_{\gb}f = f$ for all $f \in
C_0^{\infty}(M,\R)$ that
$$ \delta E_{\gb}^{\rm ret} = - \Eo{}^{\rm ret} (\delta K_{\gb})
\Eo{}^{\rm ret}\,,$$
and thus we obtain
$$ \delta F_{\gb}\varphi = \Eo\Ko\chi^{\rm ret}_-\Eo{}^{\rm ret}(\delta
K_{\gb})\Eo{}^{\rm ret} \Ko \chi^{\rm ret}_+\varphi\,.$$
Now we use the same support arguments as before to conclude that $\chi^{\rm
  ret}_-\Eo{}^{\rm adv}\delta K_{\gb} = 0$ and $\delta K_{\gb}
\Eo{}^{\rm adv}\Ko\chi^{\rm ret}_+\varphi = 0$, and hence it holds
that
$$ \delta F_{\gb}\varphi = \Eo\Ko\chi^{\rm ret}_-\Eo(\delta K_{\gb}) \Eo
\Ko \chi^{\rm ret}_+\varphi = \Eo(\delta K_{\gb})\varphi$$
for all $\varphi \in \calRo$.

Therefore, our discussion so far shows that \eqref{derivative} is
proved as soon as we have shown that, given $\gb \in G$,
\begin{eqnarray} \label{no1}
\lefteqn{\int \langle \theta, \{\Fo(x),W_{\omega}(\varphi)\}\theta
  \rangle (\delta K_{\gb}\varphi)(x)\,(-\ggo)^{1/2}(x)\,dx} \nonumber
\\
& = & -\int \langle \theta,
[T^{\mu\nu}(x),W_{\omega}(\varphi)]\theta\rangle \delta
\gb_{\mu\nu}(x) (-\ggo)^{1/2}(x)\,dx
\end{eqnarray}
holds for all $\varphi \in \calRo$ and all $\theta \in
\mathcal{V}_{\omega}$; note that $\delta K_{\gb}$ is a differential
operator on $C^{\infty}(M,\R)$ containing $\delta \gb_{\mu\nu}$. 
Due to the coordinate-independent nature of the integrals,
it is sufficient to demonstrate that \eqref{no1} holds in some
arbitrarily chosen coordinate system for $M$. We may thus choose
coordinates so that $-\ggo(x) = 1$ for all $x$. In such coordinates, 
one obtains
$$ \delta K_{\gb} = \frac{1}{2}
\gbo{}_{\mu\nu}(\partial^{\mu}(\gbo{}^{\alpha\beta}\delta
\gb_{\alpha\beta}))\partial^{\nu} -\partial^{\mu}\delta
\gb_{\mu\nu}\partial^{\nu} \quad (-\ggo = 1)\,.$$
Making also use of the fact that the $\delta \gb_{\mu\nu}$ are
compactly supported, and that $\varphi$ is a solution of the
Klein-Gordon equation \eqref{KGeqn} on $(M,\gbo)$ (for $\xi = 0$), 
one obtains after
partial integration in coordinates where $-\ggo = 1$,
\begin{eqnarray}\label{no2}
& & \ \ \int \langle
\theta,\{\Fo(x),W_{\omega}(\varphi)\}\theta\rangle (\delta 
K_{\gb}\varphi)(x)\,dx \\
& &\!\! =
\int\!\!\!\left(\langle\theta,\{\partial^{\mu}
\Fo(x),W_{\omega}(\varphi)\} 
\theta\rangle\partial^{\nu}\varphi(x) 
  -\frac{1}{2}\gbo{}^{\mu\nu}(x)
    \langle\theta,\{\partial^{\alpha}\Fo(x),
    W_{\omega}(\varphi)\}\theta\rangle\partial_{\alpha}\varphi(x)\right.
\nonumber \\ 
& & \ \ \quad \ \ + \left.\frac{1}{2}\gbo{}^{\mu\nu}(x)
    m^2\langle\theta,\{\Fo(x),W_{\omega}(\varphi)\}\theta\rangle \varphi(x)
\right)\delta \gb_{\mu\nu}(x)\,dx\,. \nonumber
\end{eqnarray}
We shall next investigate the right hand side of
\eqref{derivative}. The commutator of $W_{\omega}(\varphi)$ with the
formal, point-split expression for the bitensor $T^{\mu\nu}(x,x')$ is given by
\begin{eqnarray*}
& &
{}\hspace*{-0.7cm}\langle\theta,[T^{\mu\nu}(x,x'),W_{\omega}(\varphi)]\theta\rangle 
\ =\ \langle \theta,[\partial^{\mu}\Fo(x)\partial^{\nu}
\Fo(x'),W_{\omega}(\varphi)]\theta\rangle \\
& & {}\hspace*{-0.7cm} -
\frac{1}{2}\gbo{}^{\mu\rho}(x)Y_{\rho}{}^{\nu}(x,x')\langle\theta, 
[(\partial_{\alpha}\Fo(x)Y^{\alpha}{}_{\beta}(x,x')
\partial^{\beta}\Fo(x')
- m^2\Fo(x)\Fo(x')),W_{\omega}(\varphi)]\theta\rangle
\end{eqnarray*}
where $Y^{\nu}{}_{\alpha}(x,x')$ denotes the bitensor of parallel transport of
vectors in $T_{x'}M$ to $T_xM$. In order to be able to take the limit
$x' \to x$, one uses the relations
\begin{eqnarray*}
& [\Fo(h),W_{\omega}(\varphi)] =
i[\Fo(h),\Fco(\varphi)]W_{\omega}(\varphi) \ \ \ {\rm and} \\ 
& i[\Fo(x),\Fco(\varphi)] = - \varphi(x)\,, \quad h \in
C^{\infty}_0(M,\R),\ \varphi \in \calRo\,;
\end{eqnarray*}
the first relation holds generally in quasifree representations of the
CCR-algebra as a consequence of the Weyl-relations, and the second
relation is easily deduced from the equations
\begin{eqnarray*}
& [\Fo(h),\Fco(\varphi)] = i\sigma(Eh,\varphi) = i \int
h\,\varphi\,(-\ggo)^{1/2}\,dx\,,\\
& \langle \theta,[\Fo(h),\Fco(\varphi)]\theta\rangle = \int \langle
\theta,[\Fo(x),\Fco(\varphi)]\theta\rangle h(x)(-\ggo)^{1/2}(x)\,dx
\end{eqnarray*}
which hold for all $h \in C_0^{\infty}(M,\R)$, $\theta \in
\mathcal{V}_{\omega}$. Inserting these relations together with the
identity $[AB,C] = [A,C]B + A[B,C]$ yields for all $\theta \in
\mathcal{V}_{\omega}$
\begin{eqnarray*}
\lefteqn{\langle\theta,[T^{\mu\nu}(x,x'),W_{\omega}(\varphi)]\theta\rangle} \\
& = &- \langle
\theta,(\partial^{\mu}\Fo(x)W_{\omega}(\varphi)\partial^{\nu}\varphi(x')
+ \partial^{\mu}\varphi(x)W_{\omega}(\varphi)\partial^{\mu}\Fo(x'))
\theta\rangle\\ 
& & +
\frac{1}{2}\gbo{}^{\mu\rho}(x)Y_{\rho}{}^{\nu}\langle\theta, 
Y^{\alpha}{}_{\beta}(\partial_{\alpha}\Fo(x)W_{\omega}(\varphi)
\partial^{\beta}\varphi(x') 
+ \partial_{\alpha}\varphi(x)W_{\omega}(\varphi)
\partial^{\beta}\Fo(x'))\theta\rangle\\ 
& & - \frac{1}{2}\gbo{}^{\mu\rho}(x)Y_{\rho}{}^{\nu}m^2\langle
\theta,(\Fo(x)W_{\omega}(\varphi)\varphi(x') + 
\varphi(x)W_{\omega}(\varphi)\Fo(x'))\theta\rangle\,,
\end{eqnarray*} 
where we have abbreviated $Y_{\rho}{}^{\nu}(x,x')$ by
$Y_{\rho}{}^{\nu}$, etc.
In the last expressions, one can clearly take the limit $x' \to x$
without occurrence of any divergencies to obtain, upon observing
$\delta\gb_{\mu\nu} = \delta\gb_{\nu\mu}$,
\begin{eqnarray} \label{no3}
\lefteqn{\langle\theta,[T^{\mu\nu}(x),W_{\omega}(\varphi)]\theta\rangle
  \delta\gb_{\mu\nu}(x)} \\
& = & - \langle \theta,\{\partial^{\mu}\Fo(x),W_{\omega}(\varphi)\}
\theta\rangle\partial^{\nu}\varphi(x)\delta\gb_{\mu\nu}(x)\nonumber\\ 
& & +
\frac{1}{2}\gbo{}^{\mu\nu}(x)\langle\theta,
\{\partial_{\alpha}\Fo(x),W_{\omega}(\varphi)\}\theta\rangle 
\partial^{\alpha}\varphi(x)\delta\gb_{\mu\nu}(x)\nonumber \\
& & - \frac{1}{2}\gbo{}^{\mu\nu}(x)m^2\langle
\theta,\{\Fo(x),W_{\omega}(\varphi)\}\theta\rangle
\varphi(x)\delta\gb_{\mu\nu}(x)\,.
\nonumber 
\end{eqnarray}
Comparing \eqref{no2} and \eqref{no3}, one can see that the right hand
side and the left hand side of \eqref{no1} agree for coordinates where
$-\ggo = 1$, for all $\varphi \in \calRo$, $\theta \in
\mathcal{V}_{\omega}$, and all $\gb \in G$.  
As this is sufficient for the validity of
\eqref{derivative}, the proof is complete. {} \hfill $\Box$
${}$
\section{Wick-Polynomials} \label{Sec5}
The enlarged local algebras 
formed by the Wick polynomials defined in \cite{BFK} (with the extended
microlocal domain defined in \cite{BF}), which can be constructed in
representations of quasifree Hadamard states of the free field over
globally hyperbolic spacetimes, also satisfy the condition of local covariance.
However, the Wick-polynomials themselves are in general not locally
covariant quantum fields in the sense of Def.\ \ref{nattrans}.

This point has been taken up recently by Hollands and Wald \cite{HW},
and they have shown (among other things) that one may suitably define
Wick-polynomials of 
the free scalar field which have the property to be locally covariant
quantum fields in the sense of Def.\ \ref{nattrans}. Here, we sketch an
alternative --yet very much related-- approach to constructing such
locally covariant Wick-powers 
which emphasizes the cohomological nature of the problem.

Let, for
$(M,\gb) \in \obj(\frakM)$, $\omega_{(M,\gb)}$ be a quasifree Hadamard
state of the Klein-Gordon field on $(M,\gb)$, 
and let $:\!\Phi^n\!:_{\omega_{(M,\gb)}}$ denote the $n$-th Wick-power
 of the quantized Klein-Gordon field in the
GNS-representation of $\omega=\omega_{(M,\gb)}$, as defined in \cite{BFK,BF}.
By $\euW_{\omega}(M,\gb)$ we denote the unital $*$-algebra formed by all these
Wick-powers. When $\omega'= \omega'_{(M,\gb)}$ is another quasifree
Hadamard state, then one can show (cf.\ \cite{HW}) that there is a
$*$-isomorphism 
$\hat{\alpha}:\euW_{\omega}(M,\gb) \to \euW_{\omega'}(M,\gb)$, so up
to isomorphisms, $\euW_{\omega}(M,\gb)$ is independent of the chosen
quasifree Hadamard state. Now, to illustrate that the Wick-powers
defined with respect to a reference Hadamard state are not locally
covariant, and how they may be re-defined to become locally covariant in
terms of the solution of a cohomological problem, we will consider
in the following the case of the Wick-square.

In order that the family $\{:\!\Phi^2\!:_{\omega_{(M,\gb)}}\}$ of
Wick-squares, defined with respect to a family $\{\omega_{(M,\gb)}\}$
of quasifree Hadamard states, be a locally covariant quantum field,
 it is required that there is for any
$\psi \in \Hom_{\frakM}((M,\gb),(M',\gb'))$ a unital $*$-algebraic morphism
$\gamma_{\psi}: \euW_{\omega}(M,\gb) \to \euW_\omega(M',\gb')$ so that
$\gamma_{\psi}(:\!\Phi^2\!:_{\omega_{(M,\gb)}}) = \,
:\!\Phi^2\!:_{\omega_{(M',\gb')}} \circ \psi_*$. But actually it holds
that
$$  \tilde{\alpha}_{\psi}(:\!\Phi^2\!:_{\omega_{(M,\gb)}}\!(x)) =\,
:\!\Phi^2\!:_{\omega_{(M',\gb')}}\!(\psi(x)) +
(w^{(\omega_{(M',\gb')})}_2(\psi(x),\psi(x)) -
w^{(\omega_{(M,\gb)})}_2(x,x))$$
where we have indicated distributions by their variable entries
$(x)$, and $\tilde{\alpha}_{\psi}$ is an appropriate extension of the
$C^*$-algebraic morphism associated with the Klein-Gordon field's
functor $\euA$. Thus, the difference term on the left hand side 
 will have to vanish in order that
$\{:\!\Phi^2\!:_{\omega_{(M,\gb)}}\}$ be a locally covariant field. As
we have already indicated in Sec.\ \ref{FDSS}, this will not hold in general
(see also \cite{HW}). 

Let us indicate how this problem may be solved. If
$\omega=\omega_{(M',\gb')}$ and $\omega'=\omega'_{(M',\gb')}$ are two
quasifree Hadamard states over the spacetime $(M',\gb')$,
 then there is a smooth
function $B_{\omega,\omega'}$ on $M'$ so that $:\!\Phi^2\!:_{\omega}\!(x')
-\,:\!\Phi^2\!:_{\omega'}\!(x') \, = B_{\omega,\omega'}(x')$. These functions
satisfy the covariance condition
$$ B_{\omega\circ \tilde{\alpha}_{\psi},\omega' \circ
  \tilde{\alpha}_{\psi}}(x) = 
B_{\omega,\omega'}(\psi(x))\,, \quad x \in M\,,$$ 
for $\psi \in
\Hom_{\frakM}((M,\gb),(M',\gb'))$, and
moreover, they fulfill a cocycle condition
$$ B_{\omega,\omega'} + B_{\omega',\omega''} + B_{\omega'',\omega} =
0\,.$$
The aim is now to trivialize this cocycle while preserving its
covariance properties. In other words, we are seeking to associate
with each quasifree Hadamard state $\omega =\omega_{(M',\gb')}$ over
$(M',\gb')$ a 
smooth function $f_{\omega_{(M',\gb')}} \in C^{\infty}(M')$ so that the
resulting family of smooth functions transforms covariantly, i.e.\
$$ f_{\omega\circ \tilde{\alpha}_{\psi}}(x) = f_{\omega}(\psi(x))\,, \quad \psi
\in \Hom_{\frakM}((M,\gb),(M',\gb'))\,,$$
and trivializes the cocycle, i.e.\
$$ B_{\omega,\omega'}(x') = f_{\omega}(x') - f_{\omega'}(x')\,, \quad x'
\in M'\,,$$
for $\omega = \omega_{(M',\gb')}$, $\omega' = \omega'_{(M',\gb')}$ any
pair of quasifree Hadamard states over $(M,\gb)$. Hence we would obtain
a locally covariant Wick-square by setting
$$ :\!\Phi^2\!:_{(M,\gb)}\!(x) = \, :\!\Phi^2\!:_{\omega_{(M,\gb)}}\!(x) -
f_{\omega_{(M,\gb)}}(x) $$
for arbitrary choice a of quasifree Hadamard state
$\omega_{(M,\gb)}$ over $(M,\gb)$.

It is not too difficult to find the solution to this cohomological
problem. Recalling the definition of the Hadamard form by Kay and Wald
\cite{KW}, one finds that the diagonal values of the smooth,
non-geometrical term $H_{\omega}$ (cf.\ \eqref{HadForm}) of the
two-point function of a quasifree Hadamard state $\omega$ have the required
properties, i.e.\ a solution of the cohomological problem is provided
by defining
$$  f_{\omega}(x) = H_{\omega}(x,x) \,, \quad x \in M\,,$$
for all quasifree Hadamard states $\omega = \omega_{(M,\gb)}$ over
$(M,\gb)$.
Actually, $H_{\omega}(x,y)$ is defined off the diagonal $x=y$
only up to a $C^\infty$-function owing to the fact that the
geometrical terms $G_{\epsilon}$ are affected by the like
ambiguity. However, one can show that this ambiguity vanishes for $y
\to x$ and that, consequently, $H_{\omega}(x,x)$ is well-defined, see
the discussion in Sec.\ 5.2 of \cite{HW}.  
 
Higher order Wick-powers which are also locally covariant may then by
obtained by differentiating the generating functional
\begin{equation*}
 :\!e^{\lambda \Phi(x)}\!:_{(M,\gb)}   = e^{\frac{1}{2}\lambda^2 
f_{\omega}(x)} 
:\! e^{\lambda\Phi(x)}\!:_{\omega}
\end{equation*}
with respect to the real parameter $\lambda$, where $\omega =
\omega_{(M,\gb)}$ is any quasifree Hadamard state over $(M,\gb)$.

Finally we remark that we have only considered 
Wick-powers without derivatives. 
A discussion of Wick-powers with derivatives is contained
in a recent work by Moretti \cite{Mor}.
\\[20pt]
{\bf Acknowledgments.} We would like to thank Stefan Hollands, Bernard
Kay and Robert Wald for discussions which were stimulating for the
development of the present work.
  
\section{Appendix}
\noindent
{\bf a)} {\it Proof of statement $(\alpha)$ in the proof of Thm.\
  \ref{ThmFolia}}\\[6pt]
It is clearly sufficient to prove that $\boldsymbol{F}(\pi_{\omega
  \circ \beta}) \subset \boldsymbol{F}(\pi_{\omega} \circ \beta)$ for
all states $\omega$ on a $C^*$-algebra $\mathcal{B}$ and all
$C^*$-algebraic endomorphisms $\beta : \mathcal{A} \to \mathcal{B}$,
where $\mathcal{A}$ is another $C^*$-algebra. Consider the
GNS-representation
$(\mathcal{H}_{\omega},\pi_{\omega},\Omega_{\omega})$ of $\mathcal{B}$
corresponding to the state $\omega$. Define a new Hilbert-space
$\mathcal{H}^{\alpha}$ as the closed subspace of
$\mathcal{H}_{\omega}$ which is spanned by
$\pi_{\omega}(\alpha(\mathcal{A}))\Omega_{\omega}$. Then we may
clearly identify the GNS-representation $(\mathcal{H}_{\omega \circ
  \alpha},\pi_{\omega \circ \alpha},\Omega_{\omega \circ \alpha})$ of
$\mathcal{A}$ induced by the state $\omega \circ \alpha$ with
$(\mathcal{H}^{\alpha},\pi_{\omega}\circ \alpha,\Omega_{\omega})$
since this triple has all the properties of the GNS-triple
corresponding to $\omega \circ \alpha$, and the GNS-triple is unique
(up to unitary identifications). Hence, if $\omega' \in
\boldsymbol{F}(\pi_{\omega \circ \alpha})$, then there is a density
matrix $\rho' = \sum_j \mu_j |\phi_j\rangle\langle\phi_j|$ with unit
vectors $\phi_j \in \mathcal{H}^{\alpha}$ such that
$$  \omega'(A) = {\rm tr}(\rho'\pi_{\omega}\circ \alpha(A)) $$
holds for all $A \in \mathcal{A}$. This density matrix is then also a
density matrix on $\mathcal{H}_{\omega} \supset \mathcal{H}^{\alpha}$,
and owing to the just displayed equality, then also $\omega' \in
\boldsymbol{F}(\pi_{\omega}\circ \alpha)$ according to the definition
of the folium of a representation.
\\ \\
{\bf b)} {\it Proof of statement $(\beta)$ in the proof of Thm.\
  \ref{ThmFolia}}\\[6pt]
We quote the following result which is proved as Prop.\ 5.3.5 in
\cite{Dix}:
Let $\mathcal{B}$ be a $C^*$-algebra and $\pi$ a representation of
$\mathcal{B}$ on some Hilbert-space $\mathcal{H}$; moreover, let
$\mathcal{H}'$ be a closed subspace of $\mathcal{H}$ which is left
invariant by $\pi(\mathcal{B})$ and non-zero, and define the
subrepresentation $\pi'(B) = \pi(B)\rest \mathcal{H}'$, $B \in
\mathcal{B}$, of $\pi$ on $\mathcal{H}'$. Then $\pi$ is
quasi-equivalent to $\pi'$ if the von Neumann algebra
$\pi(\mathcal{B})''$ is a factor.

We apply this to prove statement $(\beta)$ as follows: Let $\pi$ the
identical representation of the factor $\mathcal{N}$ on the
Hilbert-space $\mathcal{H}$, and let $\pi'$ the subrepresentation
relative to $\mathcal{H}' = \mathcal{H}_{\mathcal{N}}$. According to
the quoted result, $\boldsymbol{F}(\pi) = \boldsymbol{F}(\pi')$. And
this just says that for each density matrix $\rho$ on $\mathcal{H}$
there exists a density matrix $\rho^{\mathcal{N}}$ on
$\mathcal{H}_{\mathcal{H}} = \mathcal{H}'$ so that 
$$   {\rm tr}(\rho \cdot N) = {\rm tr}(\rho\cdot \pi(N)) = {\rm
  tr}(\rho^{\mathcal{N}}\cdot \pi'(N)) = {\rm
  tr}(\rho^{\mathcal{N}}\cdot N) $$
holds for all $N \in \mathcal{N}$.

\end{document}